\documentclass[twocolumn,amsmath,amssymb,aps,prb]{revtex4-2}

\usepackage{epstopdf}
\usepackage{amsmath}
\usepackage{braket}
\usepackage{amssymb}
\usepackage{color}
\usepackage{amssymb}
\usepackage{appendix}
\usepackage{float}
\usepackage{graphicx}
\usepackage{dcolumn}
\usepackage{bm}
\usepackage{hyperref}

\hypersetup{
            colorlinks=true,
            linkcolor=blue,
            anchorcolor=blue,
            citecolor=blue}

\begin{document}


\title{Intrinsic Second-Order magnon Thermal Hall Effect}

\author{Jun-Cen Li}
\affiliation{School of Physical Sciences, University of Chinese Academy of Sciences, Beijing 100049, China}

\author{Zhen-Gang Zhu}
\email{zgzhu@ucas.ac.cn}
\affiliation{School of Physical Sciences, University of Chinese Academy of Sciences, Beijing 100049, China}
\affiliation{School of Electronic, Electrical and Communication Engineering, University of Chinese Academy of Sciences, Beijing
100049, China}

\date{\today}

\begin{abstract}
In this paper, we study the intrinsic contribution of nonlinear magnon thermal Hall Effect. We derive the intrinsic second-order thermal Hall conductivity of magnon by the thermal scalar potential (TSP) method and the thermal vector potential (TVP) method. We find that the intrinsic second-order magnon thermal Hall conductivity is related to the thermal Berry-connection polarizability (TBCP).
We apply our theory to the monolayer ferromagnetic Hexagonal lattice, and we find that the second-order magnon thermal Hall conductivity can be controlled by changing Dzyaloshinskii-Moriya strength and applying strain.
\end{abstract}

\pacs{Valid PACS appear here}
\maketitle


\section{\label{sec:level1}Introduction}
The theory of electron transport is very important for the development of new electronic devices. With the improvement of wave packet dynamics, quantum transport theory of electrons has been developed a series of developments \cite{ref1,ref2,ref3,ref4,ref5,ref6}.

For ferromagnets and antiferromagnets, the ground state of the system is usually an ordered state with ordered spin arrangement of magnetic ions. When some
perturbation deviates the spin on a lattice from the direction of the quantized axis, the spins on neighboring lattices will be deviated as well, therefore the perturbation propagates in the form of collective motion, known as spin wave, with the quasiparticle named magnon.
In particular, magnon can be excited and propagated in electronic insulators so that, in contrast to electrons, there is no energy dissipation due to the absence of joule heat \cite{ref6.1}.
Therefore, magnons are promising to replace electrons as new information carriers. To this end, it is helpful to study the transport of magnons to develop functional devices based on new strategy. In order to regulate magnon transport in materials, various proposals have been studied extensively.

In 2010, Katsura \textit{et al.} predicted the magnon thermal Hall effect theoretically \cite{ref7}.
In the same year, Onose \textit{et al.} observed the magnon thermal Hall effect in ferromagnetic material with pyrochlorite lattice structure \cite{ref8}. 
After the work of Onose \textit{et al.}, the magnon thermal Hall effect and the magnon Hall-like effect have been found in different magnetic materials \cite{ref9,ref10,ref11,ref11.1}.
Similar to the electron, magnon thermal transport is related to Berry curvature of magnon energy bands. In 2011, Matsumoto and Murakami studied the linear thermal Hall effect \cite{ref12}, and they found the relationship between linear magnon transport and Berry curvature. 
The work of Murakami \textit{et al.} initiated the study of the topology of magnon transport \cite{ref12.1,ref12.2,ref12.3,ref12.4}. 
%
In 2012, Onose \textit{et al.} measured the Hall thermal conductivity of magnon in pyrochlore ferromagnetic insulators $Ho_2V_2O_7$ and $In_2Mn_2O_7$ \cite{ref12.5}.  And they found that the experimental results are in agreement with the theory proposed by Murakami \textit{et al.} at low temperature. \cite{ref12}. 
%
In 2013, Hoogdalem \textit{et al.} demonstrated that the equilibrium magnetic texture leads to a fictitious magnetic field which can induce the magnon thermal Hall effect \cite{ref13}.
The discovery of the magnon Hall effect has led to the study of the  topological properties of magnon
\cite{ref14}.
In 2022, Kondo and  Akagi proposed the extrinsic nonlinear magnon spin Nernst effect by Boltzmann equation \cite{ref15}. In 2023, R. Mukherjee \textit{et al.} proposed the extrinsic nonlinear thermal Hall effect of magnon by Boltzmann equation \cite{ref15.1}.  
However, at present, the theory of intrinsic nonlinear transport of magnons is still lacking.

In this work, we propose an intrinsic nonlinear magnon thermal Hall effect.
Based on the work of Matsumoto and Murakami \cite{ref12}, by introducing the Berry curvature correction induced by temperature gradient, we obtain the intrinsic second-order transport of magnon.
Then we take a numerical calculation of the two-dimensional ferromagnetic model to study the effect of strain and  Dzyaloshinskii-Moriya interaction (DMI) on the intrinsic nonlinear magnon thermal Hall conductivity.
We find that
for hexagonal ferromagnet (FM) to obtain non-zero intrinsic second-order Hall thermal conductivity of magnon, it is necessary to apply strain to break the symmetry of $\mathcal{C}_3^z$. 
In addition, more effects of strain on the intrinsic magnon second-order Hall thermal conductivity are also analyzed in this work. 
%
When the first-order and second-order thermal Hall effect coexist, a non-reciprocal effect may be induced when reversing the temperature gradient. This is a kind of diode-like behavior which may be useful in future. 

\section{\label{sec:level1} Formalism}
To describe the effect of temperature gradients in a Hamiltonian formalism, Luttinger introduced a fictitious scalar field  $\Psi$, which is called the ``gravitational" potential \cite{ref16}.
It couples to energy density $h(\mathbf{r})$. To describe the effect of the temperature gradient, the Luttinger Hamiltonian reads
\begin{equation}
\hat{H}_L=\int d^3r h(\mathbf{r})\Psi(\mathbf{r}),
\end{equation}
where ``gravitational" potential satisfies
\begin{equation}
\boldsymbol{\nabla}\Psi(\mathbf{r})=\frac{\boldsymbol{\nabla} T}{T}.
\end{equation}
In the following, we refer to this method as thermal scalar potential (TSP) method.

The method of TSP often leads to unphysical divergent results as $T\to 0$. The divergent can be eliminated by the thermal vector potential (TVP) method introduced by Tatara \cite{ref17,ref19}. 
%
In this way, the Luttinger Hamiltonian can alternatively be written as
\begin{equation}
\hat{H}_L=-\int d^3r \mathbf{J}_{\epsilon}(\mathbf{r},t)\cdot\mathbf{A}_T(\mathbf{r},t),
\end{equation}
where $\mathbf{J}_{\epsilon}$ is the energy current density and $\mathbf{A}_T$ is the TVP, which satisfies
\begin{equation}
\partial_t\mathbf{A}_T(\mathbf{r},t)=\frac{\boldsymbol{\nabla} T}{T}.
\label{defAT}
\end{equation}
In the following, we refer to this method as TVP method.
In analogy to the electromagnetism, a "thermal field" $\mathbf{E}_T$ can be defined as
\begin{equation}
\mathbf{E}_T=-\frac{\boldsymbol{\nabla} T}{T}=-\partial_t\mathbf{A}_T(\mathbf{r},t)=-\boldsymbol{\nabla}\Psi(\mathbf{r}).
\label{thermalfield}
\end{equation}

In analogy with the electron system, the dynamics of magnon wave packet can be described as \cite{ref12}
\begin{equation}
\dot{\mathbf{r}}=\frac{1}{\hbar}\frac{\partial\epsilon_n(\mathbf{k})}{\partial \mathbf{k}}-\dot{\mathbf{k}}\times\boldsymbol{\Omega}_n(\mathbf{k})
\end{equation}
\begin{equation}
\hbar\dot{\mathbf{k}}=-\boldsymbol{\nabla} U(\mathbf{r}),
\end{equation}
in which
\begin{equation}
\boldsymbol{\Omega}_n(\mathbf{k})=\boldsymbol{\nabla}_{\mathbf{k}}\times\boldsymbol{\mathcal{A}}_n(\mathbf{k})=i\left\langle\frac{\partial u_n}{\partial\mathbf{k}}|\times|\frac{\partial u_n}{\partial\mathbf{k}}\right\rangle,
\end{equation}
is Berry curvature for magnons, and $\ket{u_n(\mathbf{k})}$ is Bloch state of magnon system. And $\boldsymbol{\mathcal{A}}_n(\mathbf{k})$ is intraband Berry connection which is defined as $\boldsymbol{\mathcal{A}}_n(\mathbf{k})=\bra{u_n(\mathbf{k})}i\boldsymbol{\nabla}_{\mathbf{k}}\ket{u_n(\mathbf{k})}$.  Perturbed by temperature gradient, the Bloch state thus gets a correction. This will further lead to perturbed Berry connection, Berry curvature and eventually the Hall-like transport of magnons. In the following, we use TVP method to calculate the corrected Bloch state.

\subsection{\label{sec:level2}Corrected Bloch state}
Base on the TVP method, the perturbed Hamiltonian due to the applied temperature gradient can be obtained. And the corresponding Bloch states can be calculated by time-dependent perturbation theory. Our result is only
accurate to the first order of the temperature gradient.

\subsubsection{The perturbation from temperature gradients}
For our purpose of time-dependent calculation, we follow the method in the Appendix B of Ref. \onlinecite{ref18} and introduce
\begin{equation}
\mathbf{A}_{T}=-\frac{\mathbf{E}_{T}}{\omega}\sin{\omega t},
\label{AT}
\end{equation}
which satisfies Eq. (\ref{defAT}) under the limit of $\omega\to 0$.
Because our result is only accurate to the first order of the temperature gradient, we take
\begin{equation}
\mathbf{E}_T \approx-\frac{\boldsymbol{\nabla} T}{T_0},
\label{ET}
\end{equation}
in which $T_0$ is the temperature on the origin point.

The Hamiltonian taking into account the effect of temperature gradient can be given by \cite{ref17,ref19}
\begin{equation}
\begin{split}
\hat{H}_{T}&=\hat{H}_0\left(\hat{\mathbf{p}}-\hat{H}_0\mathbf{A}_T\right)\\
&\approx\hat{H}_0-\frac{1}{2}\mathbf{A}_{T}\cdot\left[\hat{H}_0,\frac{\partial\hat{H}_0}{\partial\hat{\mathbf{p}}}\right]_+,
\end{split}
\end{equation}
where $[A,B]_{+}=AB+BA$. Then we have
\begin{equation}
\begin{split}
\hat{H}_{T}(\mathbf{k})&=e^{-i\mathbf{k}\cdot\mathbf{r}}\hat{H}_{T}e^{i\mathbf{k}\cdot\mathbf{r}}\\
&=\hat{H}_0(\mathbf{k})-\frac{1}{2}\mathbf{A}_{T}\cdot\left[\hat{H}_0(\mathbf{k}),\frac{\partial\hat{H}_0(\mathbf{k})}{\hbar\partial \mathbf{k}}\right]_+.\\
\label{HTk}
\end{split}
\end{equation}
Here, $\hat{H}_0(\mathbf{k})$ means the Hamiltonian describing the magnon without the external perturbation, and satisfies
\begin{equation}
\hat{H}_0(\mathbf{k})\ket{u_n(\mathbf{k})}=\varepsilon_n(\mathbf{k})\ket{u_n(\mathbf{k})}.
\end{equation}
And the second term of Eq. (\ref{HTk}) $\hat{H}^{\prime}(\mathbf{k})=-\frac{1}{2}\mathbf{A}_{T}\cdot\left[\hat{H}_0(\mathbf{k}),\frac{\partial\hat{H}_0(\mathbf{k})}{\hbar\partial \mathbf{k}}\right]_+$ is the perturbation term (see Appendix \ref{AA}).

\subsubsection{Time-dependent perturbation method}
In terms of time-dependent perturbation theory, the perturbed Bloch state can be derived as
\begin{equation}
\ket{\tilde{u}_n(\mathbf{k},t)}=\sum_{m}a_m(\mathbf{k},t)e^{-i\varepsilon_m(\mathbf{k})t/\hbar}\ket{u_m(\mathbf{k})}.
\label{untilde}
\end{equation}
At the initial moment, $t=0$, we have $\ket{\tilde{u}_n(\mathbf{k},0)}=\ket{u_n(\mathbf{k})}$.
So our initial condition is $a_m(\mathbf{k},0)=\delta_{nm}$. Considering the time-dependent Schr\"{o}dinger equation, the wave-function coefficients satisfy
\begin{eqnarray}
i\hbar\dot{a}_l(
\mathbf{k},t)&e^{-i\varepsilon_l(\mathbf{k})t/\hbar} &= -a_l(\mathbf{k},t)\varepsilon_l(\mathbf{k})e^{-i\varepsilon_l(\mathbf{k})t/\hbar} \notag\\
&+& \sum_{m}a_m(\mathbf{k},t)e^{-i\varepsilon_m(\mathbf{k})t/\hbar}H_{T,lm},
\label{AEQ}
\end{eqnarray}
in which $H_{T,lm}=\bra{u_l(\mathbf{k})}\hat{H}_T(\mathbf{k})\ket{u_m(\mathbf{k})}$.
We get the wave-function coefficients
\begin{equation}
a_n(\mathbf{k},t) = 1-i\frac{1}{(\hbar\omega)^2}\left(1-\cos{\omega t}\right)\varepsilon_n(\mathbf{k})
\mathbf{E}_{T}\cdot\frac{\partial\varepsilon_n(\mathbf{k})}{\partial \mathbf{k}},
\end{equation}
and
\begin{eqnarray}
& a_{l\ne n} & (\mathbf{k},t)
=\frac{\mathbf{E}_{T}}{4\hbar^2\omega}\cdot\boldsymbol{\mathcal{A}}_{ln} \left[\varepsilon_n^2(\mathbf{k})-\varepsilon_l^2(\mathbf{k})\right] \notag\\
&\times & \left[\frac{e^{i\left[\omega-\varepsilon_{nl}(\mathbf{k})/\hbar\right]t}-1}{\omega-\varepsilon_{nl}(\mathbf{k})/\hbar}
+\frac{e^{-i\left[\omega+\varepsilon_{nl}(\mathbf{k})/\hbar\right] t}-1}{\omega+\varepsilon_{nl}(\mathbf{k})/\hbar}\right],
\label{a1nen}
\end{eqnarray}
where $\varepsilon_{nl}(\mathbf{k})=\varepsilon_n(\mathbf{k})-\varepsilon_l(\mathbf{k})$, and $\boldsymbol{\mathcal{A}}_{ln}=\bra{u_l(\mathbf{k})}i\nabla_k\ket{u_n(\mathbf{k})}$ is the interband Berry connection.
After making a phase transformation, the zero frequency and long-time limit, we get the steady perturbed Bloch state
\begin{eqnarray}
\ket{\tilde{u}_n(\mathbf{k})} &=& \ket{u_n(\mathbf{k})}+\ket{u_n^{\prime}(\mathbf{k})}  \label{corrbloch}\\
&=& \ket{u_n(\mathbf{k})}-\sum_{l\ne n}\frac{\mathbf{E}_{T}}{2}\cdot\boldsymbol{\mathcal{A}}_{ln}\frac{\varepsilon_n(\mathbf{k})+\varepsilon_l(\mathbf{k})}{\varepsilon_n(\mathbf{k})-\varepsilon_l(\mathbf{k})}\ket{u_l(\mathbf{k})},
\notag
\end{eqnarray}
in which $\ket{u_n^{\prime}(\mathbf{k})}$ is the correction of the Bloch state due to the external perturbation. The details of the time-dependent perturbation method can be found in Appendix \ref{AA}. And we derive the perturbed Bloch states using TSP method and get the same result as Eq. (\ref{corrbloch}) (see Appendix \ref{AB}).

\subsection{\label{sec:citeref}The thermal Berry-connection polarizability and the correction of Berry curvature}
With the perturbed quantum state is ready, we are able to calculate
the perturbed Berry connection (the result is also accurate to the first order of the temperature gradient)
\begin{eqnarray}
\tilde{\mathcal{A}}_{n,j}(\mathbf{k}) &=& \bra{\tilde{u}_n(\mathbf{k})}i\partial_{k_j}\ket{\tilde{u}_n(\mathbf{k})} \notag\\
&\approx & \bra{u_n(\mathbf{k})}i\partial_{k_j}\ket{u_n(\mathbf{k})} +\bra{u_n(\mathbf{k})}i\partial_{k_j}\ket{u_n^{\prime}(\mathbf{k})}  \notag\\
& + & \bra{u_n^{\prime}(\mathbf{k})}i\partial_{k_j}\ket{u_n(\mathbf{k})},
\label{Anjk}
\end{eqnarray}
in which $\bra{u_n(\mathbf{k})}i\partial_{k_j}\ket{u_n^{\prime}(\mathbf{k})}+\bra{u_n^{\prime}(\mathbf{k})}i\partial_{k_j}\ket{u_n(\mathbf{k})}$ is the correction to Berry connection, we call it $\boldsymbol{\mathcal{A}}_n^{\prime}(\mathbf{k})$. From the modified Bloch state (Eq. (\ref{corrbloch})), we can write the correction to Berry connection as
\begin{eqnarray}
\mathcal{A}_{n,j}^{\prime}(\mathbf{k})
&=& \bra{u_n(\mathbf{k})}i\partial_{k_j}\ket{u_n^{\prime}(\mathbf{k})}+\bra{u_n^{\prime}(\mathbf{k})}i\partial_{k_j}\ket{u_n(\mathbf{k})} \notag\\
&=& G_{n,ji}^{t}(\mathbf{k})E_{T,i},
\label{anjk-1}
\end{eqnarray}
in which subscripts $i, j$ mean Cartesian coordinates (x, y and z).
And $E_{T,i}$ means the component of $\mathbf{E}_{T}$ in the $i$ direction.  $\mathcal{A}_{n,j}^{\prime}(\mathbf{k})$ is the component of the Berry connection correction $\boldsymbol{\mathcal{A}}_n^{\prime}$ in the $i$ direction. 
Eq. (\ref{anjk-1}) can be rewritten in a vector form as
\begin{equation}
\boldsymbol{\mathcal{A}}_{n}^{\prime}=\hat{\mathbb{G}}_n^t(\mathbf{k})\mathbf{E}_{T},
\label{vecAn}
\end{equation}
where
\begin{equation}
\hat{\mathbb{G}}_n^t(\mathbf{k})=
\left(
\begin{array}{ccc}
    G_{n,xx}^t(\mathbf{k}) & G_{n,xy}^t(\mathbf{k}) & G_{n,xz}^t(\mathbf{k})\\
    G_{n,yx}^t(\mathbf{k}) & G_{n,yy}^t(\mathbf{k}) & G_{n,yz}^t(\mathbf{k})\\
    G_{n,zx}^t(\mathbf{k}) & G_{n,zy}^t(\mathbf{k}) & G_{n,zz}^t(\mathbf{k})\\
\end{array}
\right),
\label{matrixGnt}
\end{equation}
is a second-rank  tensor,  $\hat{\mathbb{G}}_n^t$ indicates a matrix.
And we take the Einstein summation convention in this work. The element of the matrix $\hat{\mathbb{G}}_n^t$ reads
\begin{equation}
G_{n,ji}^t(\mathbf{k})=-Re\sum_{l\ne n}\left[\varepsilon_n(\mathbf{k})+\varepsilon_l(\mathbf{k})\right]\frac{\mathcal{A}_{nl,j}\mathcal{A}_{ln,i}}{\varepsilon_n(\mathbf{k})-\varepsilon_l(\mathbf{k})}.
\label{Gnjit}
\end{equation}
The second-order nonlinear intrinsic Hall effect of electrons is related to Berry-connection polarizability (BCP), and the BCP is defined as $\mathcal{A}_{n,i}^{\prime}=G_{n,ji}E_i$, in which $E_i$ means the component of the electric field in the $i$ direction \cite{ref5,ref21}. The BCP can be  expressed as
\begin{equation}
G_{n,ji}(\mathbf{k})=2eRe\sum_{l\ne n}\frac{\mathcal{A}_{nl,j}\mathcal{A}_{ln,i}}{\varepsilon_n(\mathbf{k})-\varepsilon_l(\mathbf{k})}.
\label{Gnji}
\end{equation}
Comparing Eq. (\ref{Gnjit}) to Eq. (\ref{Gnji}), we find that $-e$ in BCP is replaced by the average of the energy of two bands $(\varepsilon_n(\mathbf{k})+\varepsilon_l(\mathbf{k}))/2$ in our case.
To reflect this analogy, we introduce the name of \textit{thermal Berry-connection polarizability} (TBCP) to represent $G_{n,ji}^t$ (see Appendix \ref{AC}).
The Eq. (\ref{vecAn}) actually describes a linear response to external $\mathbf{E}_{T}$ which can be regarded as an Onsager force. And $\boldsymbol{\mathcal{A}}_{n}$ is a response which can be regarded as an Onsager flux. Thus, in this language, $\hat{\mathbb{G}}_n^t$ is a linear response coefficient. We also can explain $\hat{\mathbb{G}}_n^t$ as a kind of susceptibility which is analogy to the polarization and external electric field, i.e. $\mathbf{P}=\epsilon_{0}\hat{\mathbb{\chi}}\mathbf{E}$, $\hat{\mathbb{\chi}}$ is the electric susceptibility (generally it is a tensor).

%
%
%
Then we can get the correction to Berry curvature as
\begin{equation}
\Omega_{n,\gamma}^{\prime}(\mathbf{k})=\epsilon_{\alpha\beta\gamma}\partial_{k_{\alpha}}\mathcal{A}_{n,\beta}^{\prime}(\mathbf{k})
=\epsilon_{\alpha\beta\gamma}\partial_{k_{\alpha}}G_{n,\beta\delta}^tE_{T,\delta},
\label{Omegangamma}
\end{equation}
in which subscripts $\alpha,\beta,\gamma$ and $\delta$ are Cartesian coordinates. Eqs. (\ref{vecAn}) and (\ref{Omegangamma}) are the one of central results of the present work. 
%
In short, when the temperature gradient is present, the thermal field (standing for the temperature gradient) will disturb the Berry connection, so as to the Berry curvature of the system. However, the thermal field $\mathbf{E}_{T}$ enters into the second-order heat current density rather than the second-order thermal Hall conductivity (see Eq. (\ref{jQi2})). 

\subsection{\label{sec:citeref}Intrinsic second-order magnon thermal Hall effect}

Matsumoto and Murakami derived the linear magnon thermal Hall effect in two-dimensional system \cite{ref12}. The result is
\begin{equation}
j_{Qi}^{(1)} =-\epsilon_{ijz}T\partial_j\left(\frac{1}{T}\right)\frac{k_B^2T^2}{\hbar V}\sum_{n\mathbf{k}}\Omega_{n,z}(\mathbf{k})c_2\left(\rho_n^B\right),
\label{jQ11}
\end{equation}
where
\begin{equation}
c_2(\rho_n^B)=(1+\rho_n^B)\left(\ln{\frac{1+\rho_n^B}{\rho_n^B}}\right)^2-(\ln{\rho_n^B})^2-2\text{Li}_2(-\rho_n^B),
\end{equation}
in which $\rho_n^B$ is the Bose-Einstein distribution $\rho_n^B=\frac{1}{e^{[\varepsilon_n(\mathbf{k})-\mu]/k_BT}-1}$, and  $\rm{Li}_s(z)$ is  the polylogarithm function defined as $\rm{Li}_s(z)=\sum_{n=1}^{\infty}\frac{z^n}{n^s}$.
In Eq. (\ref{jQ11}), $j_{Qi}^{(1)}$ is heat current density in $i$ direction related to the first order temperature gradient. The correction of Berry curvature (Eq. (\ref{Omegangamma})) induces a correction to heat current density as
\begin{eqnarray}
j_{Qi}^{\text{corrected}} &=& -\epsilon_{ijz}T \partial_j(1/T) \frac{k_B^2T^2}{\hbar V}\sum_{n\mathbf{k}}\Omega_{n,z}^{\prime}(\mathbf{k})c_2(\rho_n^B) \notag\\
&=& -\epsilon_{ijz}T \partial_j(1/T) \frac{k_B^2T^2}{\hbar V} \notag\\
&\times & \sum_{n\mathbf{k}}\left[\epsilon_{\alpha\beta z}\partial_{k_{\alpha}}G_{n,\beta\delta}^t(\mathbf{k})E_{T,\delta}\right]c_2(\rho_n^B).
\label{jQicorrected}
\label{jcor}
\end{eqnarray}
From Eq. (\ref{jQicorrected}) and keep to the second-order to the temperature gradient, we get the intrinsic second-order Hall heat current of magnon as (see Appendix \ref{HeatTBCP})
\begin{equation}
j_{Qi}^{(2)} =-\kappa_{ij\delta}(\partial_jT)(\partial_{\delta}T),
\label{jQi2}
\end{equation}
where the second-order magnon thermal Hall conductivity is introduced as 
\begin{equation}
\kappa_{ij\delta}
=-\frac{k_B^2}{\hbar V}\sum_{n\mathbf{k}}c_2(\rho_n^B)\left[\partial_{k_{j}}G_{n, i\delta}^t(\mathbf{k})-\partial_{k_{i}}G_{n, j\delta}^t(\mathbf{k})\right],
\label{kappa}
\end{equation}
in which $i,j,\delta=x,y$ (see Appendix \ref{HeatTBCP}). 
In the work of Liu \textit{et. al.}, the conductivity of the intrinsic nonlinear Hall effect is expressed as \cite{ref21}
\begin{equation}
\chi_{ij\delta}
=\sum_{n\mathbf{k}}\rho_{n}^F\left(\partial_{k_j}G_{n,i\delta}-\partial_{k_i}G_{n,j\delta}\right),
\end{equation}
which enters into the charge Hall current density as $j_i=\chi_{ij\delta}E_jE_{\delta}$.
Compared to the second-order charge Hall conductivity in electron system, the second-order magnon intrinsic thermal Hall conductivity can be obtained by replacing the Fermi-Dirac distribution $\rho_{n}^F$ with $\frac{k_B^2}{\hbar V}c_2(\rho_n^B)$, and replacing BCP with TBCP. These replacements reflect a logical parallelism between charge and thermal transport.

There are 8 elements for $\kappa_{ij\delta}$ ($i, j, \delta = x, y$). From Eq. (\ref{kappa}), we have
\begin{equation}
  \kappa_{ij\delta}=-\kappa_{ji\delta},
  \label{antisymmetry}
\end{equation}
so we get that $\kappa_{ii\delta}=0$ ($i, \delta = x, y$).
Therefore, there are only 4 nonzero elements for the second-order magnon thermal Hall conductivity, i.e. $\kappa_{xyx}$, $\kappa_{xyy}$, $\kappa_{yxx}$, and $\kappa_{yxy}$.
Due to Eq. (\ref{antisymmetry}), we have $\kappa_{yxx} = -\kappa_{xyx}$, $\kappa_{yxy} = -\kappa_{xyy}$.
So there are only 2 independent elements for $\kappa_{ij\delta}$, and we can take $\kappa_{xyx}$ and $\kappa_{xyy}$ as the independent elements.
Then we have

\begin{eqnarray}
\kappa_{xy\delta} &=& -\frac{k_B^2}{\hbar V}\sum_{n\mathbf{k}}c_2(\rho_n^B)\left[\partial_{k_{y}}G_{n, x\delta}^t(\mathbf{k})-\partial_{k_{x}}G_{n, y\delta}^t(\mathbf{k})\right] \notag\\
&=& -\frac{k_B^2}{\hbar V}\sum_{n\mathbf{k}}c_2(\rho_n^B)\left[\partial_{k_{y}}G_{n,\delta x}^t(\mathbf{k})-\partial_{k_{x}}G_{n,\delta y}^t(\mathbf{k})\right] \notag\\
&=& \frac{k_B^2}{\hbar V}\sum_{n\mathbf{k}}c_2(\rho_n^B)\left[\boldsymbol{\nabla}_k\times\mathbf{G}_{n,\delta}^t(\mathbf{k})\right]\cdot\mathbf{E}_z,
\label{kappaxydelta}
\end{eqnarray}
in which $\mathbf{G}^{t}_{n,\delta}(\mathbf{k})$ is a defined vector which is call TBCP vector, and its components are given by
\begin{equation}
\mathbf{G}_{n,\delta}^t(\mathbf{k})=
\left(
\begin{array}{ccc}
G_{n,\delta x}^t(\mathbf{k}),  G_{n,\delta y}^t(\mathbf{k}),  G_{n,\delta z}^t(\mathbf{k})
\end{array}
\right),
\label{TBCPV}
\end{equation}
which is actually the $\delta$-th row of the matrix $\hat{\mathbb{G}}_n^t$.

\begin{figure}[tb]
    \centering
    \includegraphics[width=1\columnwidth]{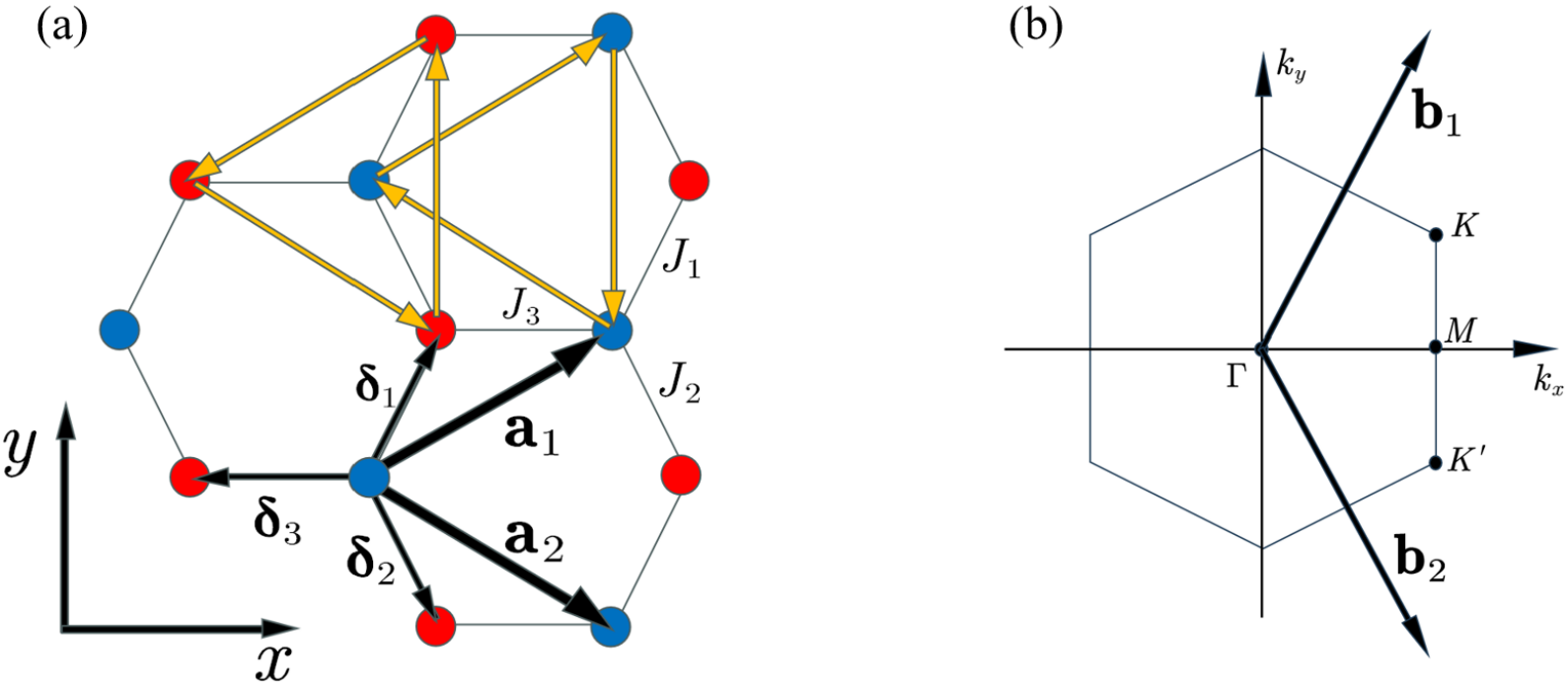}
    \caption{(a) Ferromagnetic Hexagonal lattice in real space.
    $\mathbf{a}_1$ and $\mathbf{a}_2$ are basis vectors in real space. The DM vectors are parallel to the z axis $\mathbf{D}_{ij}=D\nu_{ij}\mathbf{E}_z$, in which $\mathbf{E}_z$ represents the unit vector pointing in the positive direction of the z axis. As shown in the figure, $\nu_{ij}=1$ along the orange arrows, in which $i$ and $j$ represent the next-nearest lattice points. (b) Unit cell in the reciprocal space.}
    \label{fig:lattice}
    \end{figure}

The independent elements form a vector $\boldsymbol{\kappa}^{(2)}$, which is named as the second-order thermal Hall conductivity vector (2THCV) and defined as
\begin{equation}
\kappa_{\delta}^{(2)}
=\epsilon_{abz}\kappa_{ab\delta}/2
\end{equation}
The two independent components of this vector compose the $j_{Qx}^{(2)}$, i.e. $j_{Qx}^{(2)}=-\kappa_{xyx}(\partial_{y}T)(\partial_{x}T)-\kappa_{xyy}(\partial_{y}T)^{2}$, and $j_{Qy}^{(2)}=-\kappa_{yxx}(\partial_{x}T)^{2} -\kappa_{yxy} (\partial_{x}T)(\partial_{y}T)$.
%
%

\section{\label{sec:level1} Model Calculation}
\subsection{\label{sec:citeref}Model}

In Fig. \ref{fig:lattice}, we take ferromagnet (FM) as our models to study the intrinsic second-order thermal Hall conductivity of magnon. Here, $\mathbf{a}_1$ and $\mathbf{a}_2$ are the basis vectors for real space lattice. The Hamiltonian is given by
\begin{eqnarray}
\hat{H} &=& \sum_{\langle ij\rangle}J_{ij}\hat{\mathbf{S}}_i\cdot\hat{\mathbf{S}}_j+\sum_{\ll ij\gg }\mathbf{D}_{ij}\cdot\left(\hat{\mathbf{S}}_i\times\hat{\mathbf{S}}_j\right) \notag\\
& + & g_{J}\mu_B\sum_i\hat{\mathbf{S}}_i\cdot\mathbf{B}+ \mathcal{K}\sum_i\hat{S}_{i}^{z2}.
\label{model-study}
\end{eqnarray}

The first term is Heisenberg interaction, the second term is the DM interaction, the third term is the Zeeman interaction, and the fourth term is the easy-axis anisotropy. 
%
Eq. (\ref{model-study}) formally contains the same terms as the Hamiltonian in the study about rectification of the spin Seebeck current by L. Chotorlishvili \textit{et al.} \cite{ref21.01}. However, Eq. (\ref{model-study}) describes a two-dimensional hexagonal ferromagnetic lattice, and the Hamiltonian in the work of L. Chotorlishvili \textit{et al.} describes the antiferromagnetic chain and two ferromagnetic chains with antiferromagnetic coupling \cite{ref21.01}. 

For the symmetry of $\mathcal{C}_3^z$ and $\mathcal{P}$, 2THCV satisfies
\begin{equation}
\boldsymbol{\kappa}^{(2)}=\hat{O}\boldsymbol{\kappa}^{(2)}.
\end{equation}
Here $\hat{O}$ is $\mathcal{C}_3^z$ and $\mathcal{P}$, in which $\mathcal{P}$ means $x\to-x, y\to-y$. Therefore the intrinsic second-order magnon thermal Hall conductivity $\kappa_{xyy}=0$ under the symmetry of $\mathcal{C}_3^z$ and $\mathcal{P}$.
To break the symmetry of $\mathcal{P}$, we need to apply a periodic magnetic field in the ferromagnetic Hexagonal lattice so that the red and blue lattice points are subjected to opposite magnetic field.
To break the symmetry of $\mathcal{C}_3^z$, we need to apply an uniaxial strain.
Next, we study the effects of uniaxial strain and DM interaction on the intrinsic second-order magnon thermal Hall conductivity $\kappa_{xyy}$.

\begin{figure}[tb]
    \centering
    \includegraphics[width=1\columnwidth]{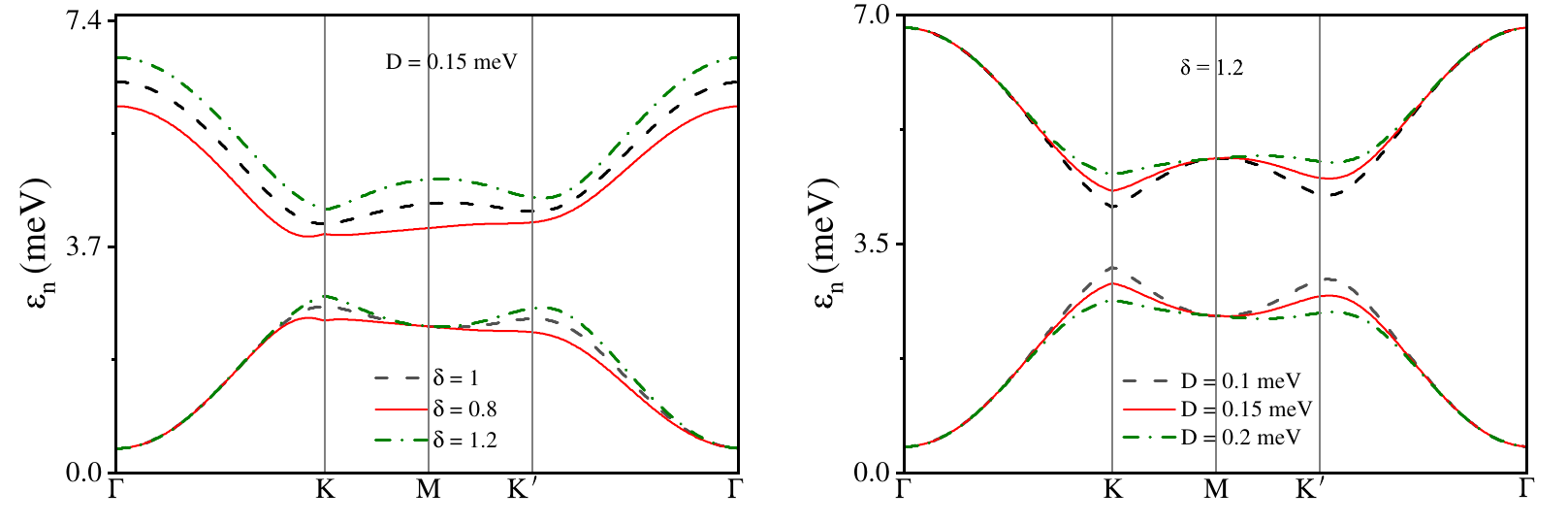}
    \caption{(a) Magnon bands with different strain, in which $D = 0.15$ meV. (b) Magnon bands with different DM srtength, in which $\delta = 1.2$.}
    \label{fig:band}
    \end{figure}

The Hamiltonian can be written as
\begin{equation}
\hat{H}
=\sum_{\mathbf{k}}\hat{\Phi}^{\dag}(\mathbf{k})\mathcal{H}(\mathbf{k})\hat{\Phi}(\mathbf{k})
\end{equation}
in which 
\begin{equation}
\hat{\Phi}(\mathbf{k})=(\hat{a}_{\mathbf{k}},\hat{b}_{\mathbf{k}})^{T}, \notag \\
\end{equation}
\begin{equation}
\mathcal{H}(\mathbf{k})=\begin{pmatrix}
\Delta_{BSa}+\Delta(\mathbf{k}) & 3JS\gamma_{\mathbf{k}}\\
3JS\gamma_{-\mathbf{k}} & \Delta_{BSb}-\Delta(\mathbf{k})\\
\end{pmatrix}.
\label{Hamiltonian}
\end{equation}
Here, $\Delta_{BSa}=-3JS-g_J\mu_BB-2\mathcal{K}S$, $\Delta_{BSb}=-3JS+g_J\mu_BB-2\mathcal{K}S$, $J=\left(J_1+J_2+J_3\right)/3$, $\gamma_{\mathbf{k}}=\frac{1}{3J}\sum_iJ_ie^{i\mathbf{k}\cdot\mathbf{\delta_i}}$, and $\Delta(\mathbf{k})=2SD\left[\sin{\frac{1}{2}(k_{y}-k_{x}\sqrt{3})}-\sin{k_{y}}+\sin{\frac{1}{2}(k_{y}+k_{x}\sqrt{3})}\right]$ (see Appendix \ref{AE}).

\begin{figure}[tb]
    \centering
    \includegraphics[width=1\columnwidth]{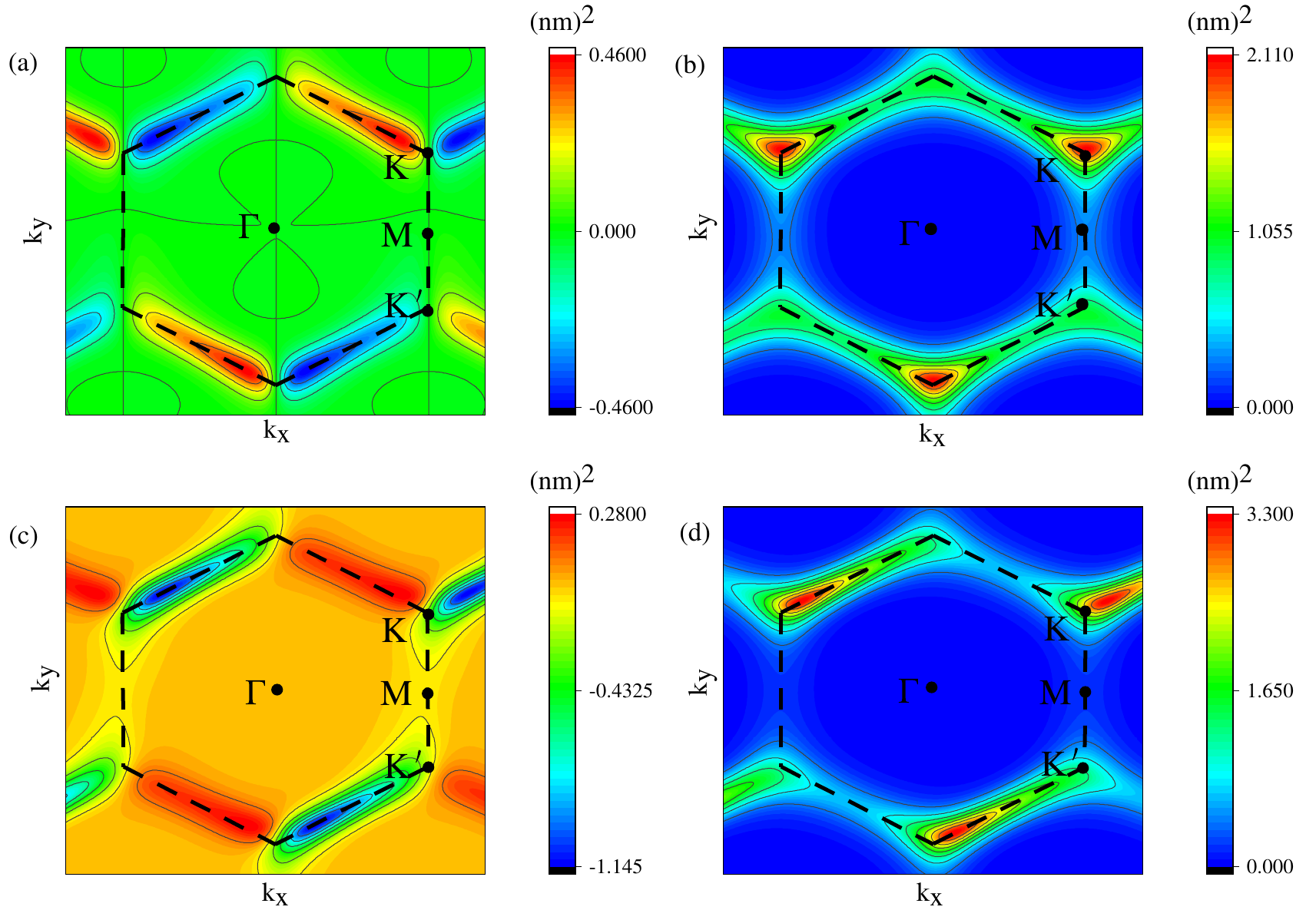}
    \caption{TBCP of lower band with different strain, in which $D = 0.15$ meV. (a) $G_{0,xy}^t(\mathbf{k})$ without strain. (b) $G_{0,yy}^t(\mathbf{k})$ without strain. (c) $G_{0,xy}^t(\mathbf{k})$ with strain of $\delta = 1.2$. (d) $G_{0,yy}^t(\mathbf{k})$ with strain of $\delta = 1.2$.}
    \label{fig:Gstrain}
    \end{figure}

\subsection{\label{sec:citeref}Numerical result}
We investigate the effect of strain and DM strength on the intrinsic second-order thermal Hall conductivity of magnons in ferromagnetic Hexagonal lattice.

To describe the effect of uniaxial strain along the direction of $\boldsymbol{\delta}_2$, we follow the method in Ref. \onlinecite{ref15} and Ref. \onlinecite{ref21.1} in which $J_{1}=J_{3}=-1$ meV is kept and $J_2$ running as a parameter. We use $J_2=J_2^0\delta$ where $J_2^0=-1$  meV and $\delta$ takes into account the effect of strain on changing of $J_{2}$. 
Here, $J_2$ is the Heisenberg spin exchange interaction with the uniaxial strain applied along the direction of  $\boldsymbol{\delta}_2$ and $J_2^0$ is the Heisenberg spin exchange interaction without the strain applied. So $\delta=J_2/J_2^0$ describes the change proportion of $J_2$ under uniaxial strain. Phenomenally speaking, we think that when the distance between atoms is reduced by the uniaxial strain, the overlap of electron cloud between atoms increases, which results in an increase of $|J_2|$ ($\delta > 1$). When the distance between atoms is increased by the uniaxial strain, the overlap of electron cloud between atoms decreases, which results in an decrease of $|J_2|$ ($\delta < 1$). So we use $\delta$ to represent the uniaxial strain along the direction of $\boldsymbol{\delta}_2$. And we make the assumption that only the Heisenberg spin exchange interaction along $\boldsymbol{\delta}_2$ changes, without lattice deformation \cite{ref21.1}. 
%
And in our work, we take $g_J\mu_BB = -0.1$ meV, and $\mathcal{K} = - 0.2$ meV.

In Fig. \ref{fig:band} (a), we plot the magnon spectrums with different stain.
For the upper bands, the strain of $\delta=0.8$ decreases the energy of magnons, and the strain of $\delta=1.2$ increases the energy of magnons globally.
For the lower bands, the strain of $\delta=0.8$ decreases the energy of magnons around point $K$ and point $K^{\prime}$, the strain of $\delta=1.2$ increases the energy of magnons around point $K$ and point $K^{\prime}$.
In Fig. \ref{fig:band} (b), we plot the magnon spectrums with different DM strength.
The increasing of DM strength increases the energy difference $\Delta\varepsilon$ around points $K$ and $K^{\prime}$.

\begin{figure}[t]
    \centering
    \includegraphics[width=1\columnwidth]{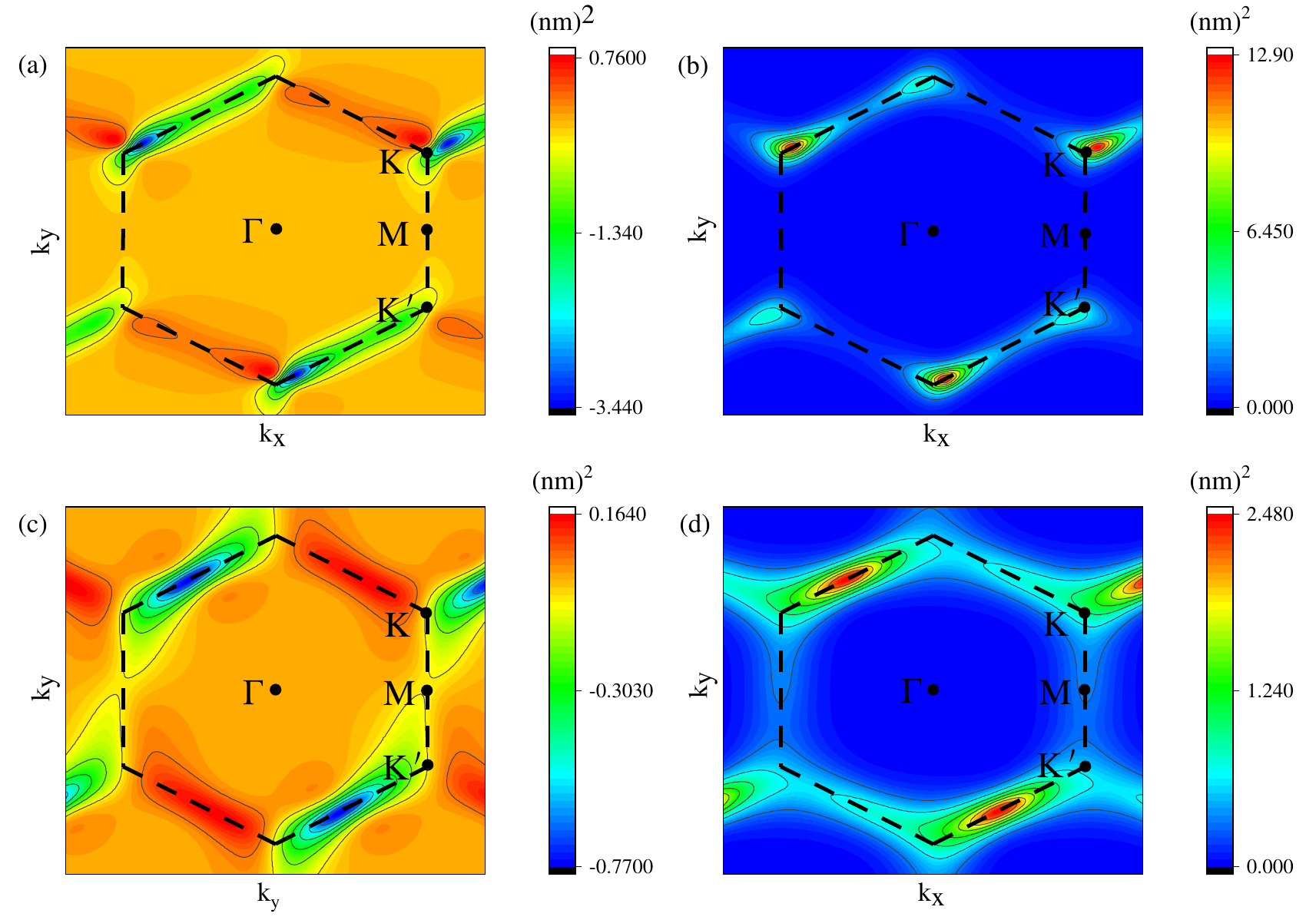}
    \caption{TBCP of lower band with different DM strength, in which $\delta = 1.2$. (a) $G_{0,xy}^t(\mathbf{k})$ with $D = 0.1$ meV. (b) $G_{0,yy}^t(\mathbf{k})$ with $D = 0.1$ meV. (c) $G_{0,xy}^t(\mathbf{k})$ with $D = 0.2$ meV. (d) $G_{0,yy}^t(\mathbf{k})$ with $D = 0.2$ meV. }
    \label{fig:GDMI}
    \end{figure}

\begin{figure}[t]
	\centering
	\includegraphics[width=1\columnwidth]{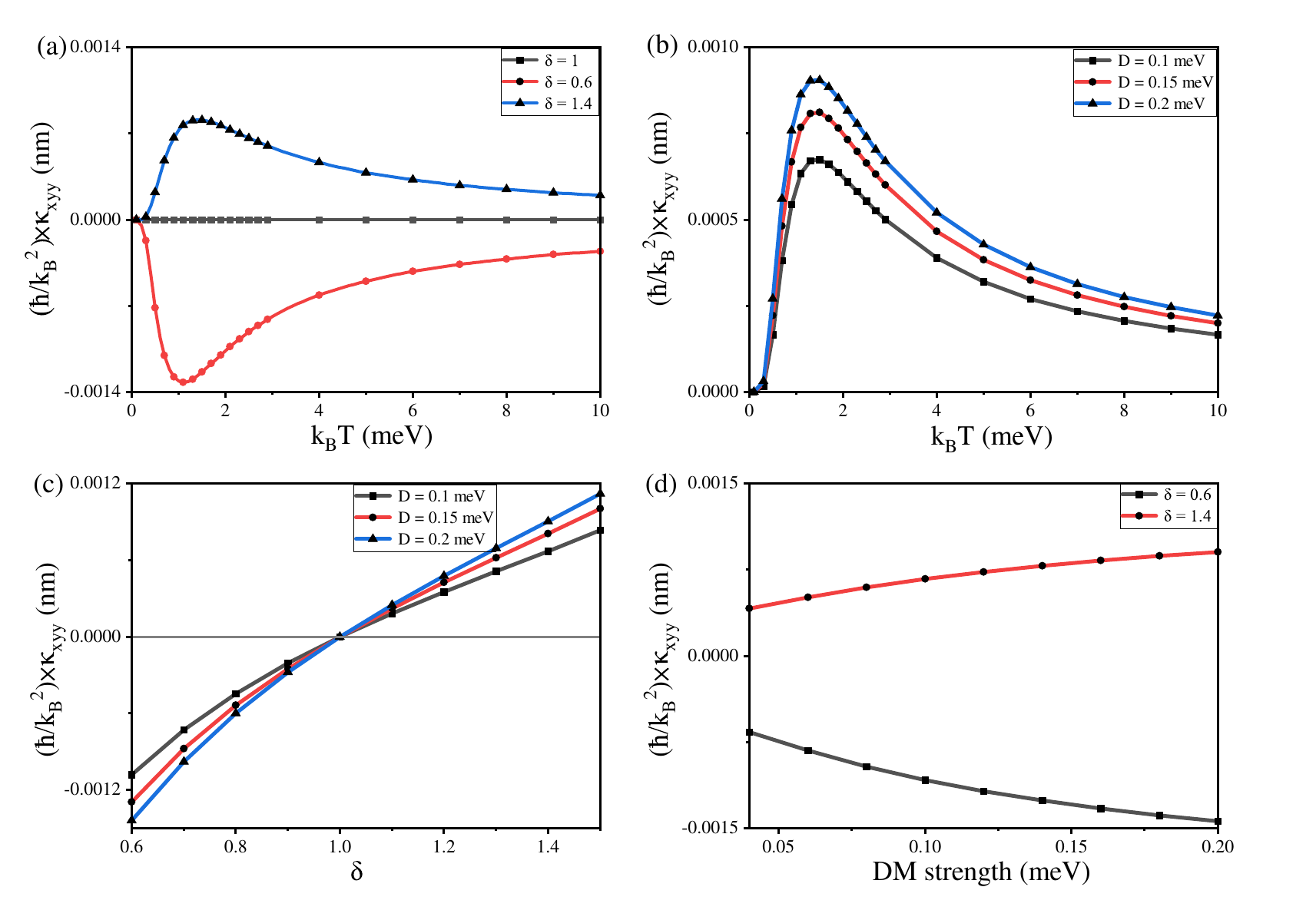}
	\caption{(a) $\kappa_{xyy}$ as a function of temperature under different strain, in which $D = 0.15$ meV. (b) $\kappa_{xyy}$ as a function of temperature under different DM strength, in which $\delta = 1.2$. (c) $\kappa_{xyy}$ as a function of strain with $k_BT = 1.3$ meV. (d) $\kappa_{xyy}$ as a function of DM strength with $k_BT = 1.3$ meV.}
	\label{fig:kappa}
\end{figure}
\begin{figure}[t]
	\centering
	\includegraphics[width=0.7\columnwidth]{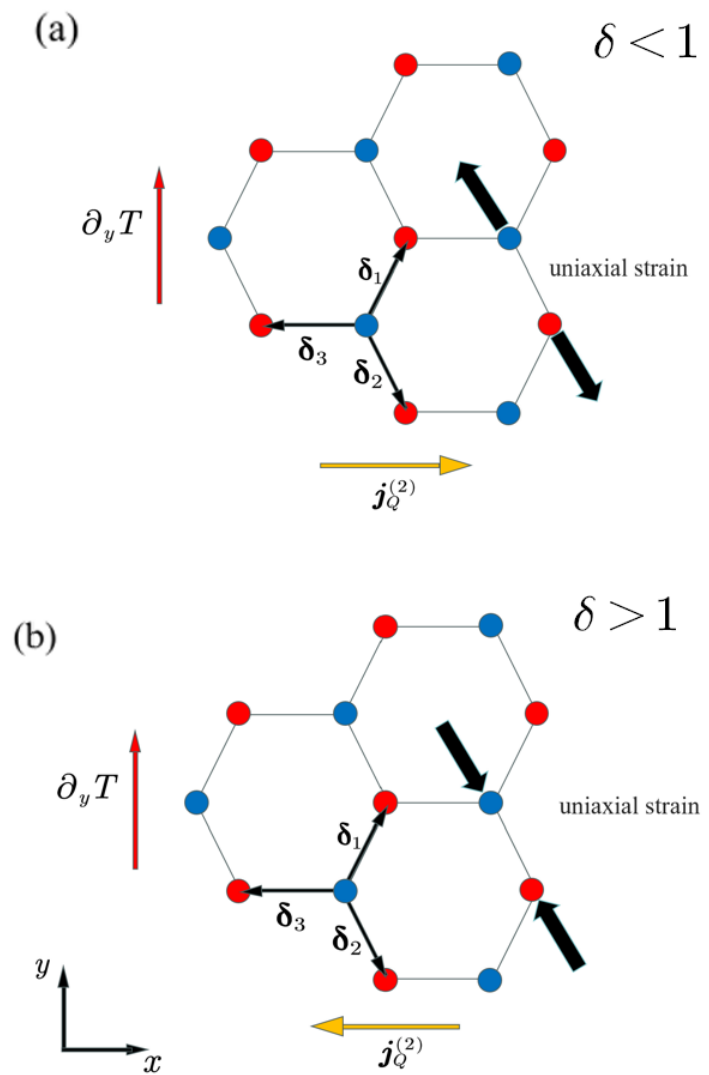}
	\caption{The direction of magnon intrinsic second-order Hall heat current under the temperature gradient along the direction of y-axis. (a) An uniaxial strain in the tensile direction ($\delta<1$) is applied to the system. (b) An uniaxial strain in the compression direction ($\delta>1$) is applied to the system.}
	\label{fig:jstrain}
\end{figure}

In Fig. \ref{fig:Gstrain}, we plot TBCP of lower band with different strain. Without strain, TBCP satisfies $G_{0,yy}^t(-k_x,k_y)=G_{0,yy}^t(k_x,k_y)$ and $G_{0,xy}^t(-k_x,k_y)=-G_{0,xy}^t(k_x,k_y)$.
This property makes $\left[\partial_{k_{y}}G_{0, xy}^t(\mathbf{k})-\partial_{k_{x}}G_{0, yy}^t(\mathbf{k})\right]$ in Eq. ($\ref{kappa}$) an odd function along the $k_x$ direction. Applying a strain breaks this property.

In Fig. \ref{fig:GDMI}, we plot TBCP of lower band with different DM strength.
As Eq. ($\ref{Gnjit}$), the denominator of TBCP is band difference $\triangle\varepsilon$, so the changing of TBCP is roughly caused by the changing of $\triangle\varepsilon$. We take $G_{0,yy}^t(\mathbf{k})$ for an illustration, as DM strength $D$ increases, $\triangle\varepsilon$ increases at point $K$ and point $K^{\prime}$, so the magnitude of TBCP decreases at point $K$ and point $K^{\prime}$.

Then we study the effects of strain and changing DM strength on $\kappa_{xyy}$.
In Fig. \ref{fig:kappa} (a) and (b), we plot $\kappa_{xyy}$ curve with respect to temperature under different strain and different DM strength respectively.
As the temperature rises, $\kappa_{xyy}$ increases rapidly at low temperature, and decreases after reaching a maximum value.
There is no $\kappa_{xyy}$ without strain. $\kappa_{xyy}$ with strain of $\delta=0.8$ and $\delta=1.2$ have different sign. And the $\kappa_{xyy}$ curve increases with the increasing of DM strength.
As shown in Fig. \ref{fig:kappa} (c), we can reverse the direction of intrinsic second-order magnon Hall heat current by changing the direction of the strain. And we can increase the intrinsic second-order magnon Hall thermal conductivity by increasing the strength of strain.
As shown in Fig. \ref{fig:kappa} (d), the intrinsic second-order magnon thermal Hall conductivity increases with the increasing of DM strength.

In Fig. \ref{fig:jstrain}, we plot a schematic diagram of the direction of intrinsic second-order Hall heat current of magnon when a temperature gradient along the direction of y-axis is applied. When we apply the tensile uniaxial strain ($\delta<1$) along the $\boldsymbol{\delta}_2$ direction, the second-order intrinsic Hall heat current of magnon flows in the positive direction of x-axis. When we apply the compression uniaxial strain ($\delta>1$) along the $\boldsymbol{\delta}_2$ direction, the second-order intrinsic Hall heat current of magnon flows in the negative direction of x-axis. 

\section{\label{sec:level1} conclusion}
In conclusion, we have proposed the intrinsic second-order magnon thermal Hall effect by perturbation theory.
And we have investigated the intrinsic second-order magnon thermal Hall conductivity in the monolayer ferromagnetic Hexagonal lattice. Then we have found that the intrinsic second-order magnon thermal Hall conductivity can be controlled by applying strain and changing DM strength. The sign and the magnitude of intrinsic nonlinear magnon thermal Hall conductivity can be controlled altering the strain, and DM strength.

When the second version of the present manuscript is being prepared, we note a similar work from Harsh Varshney \textit{et al.} in which they calculated the density matrix modification of temperature gradient, then calculated the thermodynamic average value of magnon energy multiplied by velocity and the energy magnetization current through the density matrix \cite{ref22}. The intrinsic part of heat current has been extracted under the dilute impurity limit (DIL).
In our work, we perform extra symmetry analysis in two-dimensional hexagonal ferromagnetic systems.
We find that for existence of the non-zero intrinsic nonlinear Hall thermal conductivity of magnon in hexagonal ferromagnet, it is necessary to apply a strain to break the symmetry of $\mathcal{C}_3^z$. However, in the model calculation by Harsh Varshney et al., intrinsic nonlinear magnon thermal conductivity can be obtained in non-strain hexagonal ferromagnet model, which is different to our results. In addition, we propose the notion of thermal Berry-connection polarizability and work out compact formalism with transparent physical picture, which is out of the scale of Ref.  \onlinecite{ref22}. Moreover, we study the effects of strain on the second order intrinsic nonlinear thermal Hall transport. 
%
Since the aim of present work is to illustrate the existence of such an intrinsic second-order magnon thermal Hall effect, the magnon-magnon interaction has not been included and is deserved a further study. 

\section{Acknowledgements}
This work is supported by the National Key R\&D Program of China (Grant No. No. 2022YFA1402802).
It is also supported in part by the NSFC (Grants No. 11974348 and No. 11834014), and the Strategic Priority
Research Program of CAS (Grants No. XDB28000000, and No. XDB33000000). Z.G.Z. is supported in part by
the Training Program of Major Research plan of the National Natural Science Foundation of China (Grant No.
92165105), and CAS Project for Young Scientists in Basic Research Grant No. YSBR-057.

\appendix
\begin{widetext}

\section{The detail of the TVP perturbation method} \label{AA}

From the Eq. (\ref{HTk}) in the main text, we can get
\begin{equation}
\begin{split}
\hat{H}_T(\mathbf{k})&=e^{-i\mathbf{k}\cdot\mathbf{r}}\hat{H}_Te^{i\mathbf{k}\cdot\mathbf{r}}\\
&=e^{-i\mathbf{k}\cdot\mathbf{r}}\left\{\hat{H}_0-\frac{1}{2}\mathbf{A}_{T}\cdot\left[\hat{H}_0,\frac{\partial\hat{H}_0}{\partial\hat{\mathbf{p}}}\right]_+\right\}e^{i\mathbf{k}\cdot\mathbf{r}}\\
&=\hat{H}_0(\mathbf{k})-\frac{1}{2}\mathbf{A}_{T}\cdot e^{-i\mathbf{k}\cdot\mathbf{r}}\left[\hat{H}_0,\frac{\partial\hat{H}_0}{\partial\hat{\mathbf{p}}}\right]_+e^{i\mathbf{k}\cdot\mathbf{r}}\\
&=\hat{H}_0(\mathbf{k})-\frac{1}{2}\mathbf{A}_{T}\cdot\left[\hat{H}_0(\mathbf{k}),e^{-i\mathbf{k}\cdot\mathbf{r}}\frac{\partial\hat{H}_0}{\partial\hat{\mathbf{p}}}e^{i\mathbf{k}\cdot\mathbf{r}}\right]_+\\
&=\hat{H}_0(\mathbf{k})-\frac{1}{2}\mathbf{A}_{T}\left[\hat{H}_0(\mathbf{k}),\frac{\partial\hat{H}_0(\mathbf{k})}{\hbar\partial \mathbf{k}}\right]_+,\\
\end{split}
\end{equation}
As the Eq. (\ref{untilde}) in the main text, the corrected quantum state is
\begin{equation}
\begin{split}
\ket{\tilde{u}_n(\mathbf{k},t)}=\sum_{m}a_m(\mathbf{k},t)e^{-i\varepsilon_m(\mathbf{k})t/\hbar}\ket{u_m(\mathbf{k})}.
\label{Acorrbloch}
\end{split}
\end{equation}
Substituting Eq. (\ref{Acorrbloch}) into the time-dependent Schr\"{o}dinger equation gives us
\begin{equation}
\begin{split}
i\hbar\partial_t\left[\sum_{m}a_m(\mathbf{k},t)e^{-i\varepsilon_m(\mathbf{k})t/\hbar}\ket{u_m(\mathbf{k})}\right]=\hat{H}_T(\mathbf{k})\left[\sum_{m}a_m(\mathbf{k},t)e^{-i\varepsilon_m(\mathbf{k})t/\hbar}\ket{u_m(\mathbf{k})}\right].
\label{TDSQ}
\end{split}
\end{equation}
We take the inner product $\bra{u_l(\mathbf{k})}$ on Eq. (\ref{TDSQ}), then we can know that the wave-function coefficients satisfy
\begin{equation}
\begin{split}
i\hbar\dot{a}_l(\mathbf{k},t)e^{-i\varepsilon_l(\mathbf{k})t/\hbar}=\sum_{m}a_m(\mathbf{k},t)e^{-i\varepsilon_m(\mathbf{k})t/\hbar}H_{T,lm}-a_l(\mathbf{k},t)\epsilon_l(\mathbf{k})e^{-i\varepsilon_l(\mathbf{k})t/\hbar},
\label{aEQ}
\end{split}
\end{equation}
as Eq. (\ref{AEQ}) in the main text.
To calculate the wave-function coefficients, we need to derive the matrix element of Hamiltonian $\hat{H}_{T,lm}$
\begin{equation}
\begin{split}
H_{T,lm}&=\bra{u_l(\mathbf{k})}\hat{H}_T(\mathbf{k})\ket{u_m(\mathbf{k})}=\bra{u_l(\mathbf{k})}\left[\hat{H}_0(\mathbf{k})+\hat{H}^{\prime}(\mathbf{k})\right]\ket{u_m(\mathbf{k})}\\
&=\varepsilon_m(\mathbf{k})\delta_{lm}+\bra{u_l(\mathbf{k})}\hat{H}^{\prime}(\mathbf{k})\ket{u_m(\mathbf{k})}\\
&=\varepsilon_m(\mathbf{k})\delta_{lm}-\frac{1}{2}\mathbf{A}_{T}\cdot\bra{u_l(\mathbf{k})}\left[\hat{H}_0(\mathbf{k}),\frac{\partial\hat{H}_0(\mathbf{k})}{\hbar\partial \mathbf{k}}\right]_+\ket{u_m(\mathbf{k})}\\
&=\varepsilon_m(\mathbf{k})\delta_{lm}-\frac{1}{2}\mathbf{A}_{T}\cdot\left[\varepsilon_l(\mathbf{k})+\varepsilon_m(\mathbf{k})\right]\bra{u_l(\mathbf{k})}\frac{\partial\hat{H}_0(\mathbf{k})}{\hbar\partial \mathbf{k}}\ket{u_m(\mathbf{k})}\\
&=\varepsilon_m(\mathbf{k})\delta_{lm}-\frac{1}{2\hbar}\mathbf{A}_{T}\cdot\left[\varepsilon_l(\mathbf{k})+\varepsilon_m(\mathbf{k})\right]\left[\bra{u_l(\mathbf{k})}\frac{\partial}{\partial \mathbf{k}}\left(\hat{H}_0(\mathbf{k})\ket{u_m(\mathbf{k})}\right)-\bra{u_l(\mathbf{k})}\hat{H}_0(\mathbf{k})\frac{\partial}{\partial \mathbf{k}}\ket{u_m(\mathbf{k})}\right]\\
&=\varepsilon_m(\mathbf{k})\delta_{lm}-\frac{1}{2i\hbar}\mathbf{A}_{T}\cdot\left[\varepsilon_l(\mathbf{k})+\varepsilon_m(\mathbf{k})\right]\left[\bra{u_l(\mathbf{k})}i\frac{\partial}{\partial \mathbf{k}}\left(\varepsilon_m(\mathbf{k})\ket{u_m(\mathbf{k})}\right)-\varepsilon_l(\mathbf{k})\bra{u_l(\mathbf{k})}i\frac{\partial}{\partial \mathbf{k}}\ket{u_m(\mathbf{k})}\right]\\
&=\varepsilon_m(\mathbf{k})\delta_{lm}-\frac{1}{2i\hbar}\mathbf{A}_{T}\cdot\left[\varepsilon_l(\mathbf{k})+\varepsilon_m(\mathbf{k})\right]\left[i\frac{\partial\varepsilon_m(\mathbf{k})}{\partial \mathbf{k}}\delta_{lm}+\varepsilon_m(\mathbf{k})\boldsymbol{\mathcal{A}}_{lm}-\varepsilon_l(\mathbf{k})\boldsymbol{\mathcal{A}}_{lm}\right],\\
\end{split}
\end{equation}
in other words,
\begin{equation}
\begin{split}
H_{T,ll}
&=\varepsilon_l(\mathbf{k})-\varepsilon_l(\mathbf{k})\mathbf{A}_{T}\cdot\frac{\partial\varepsilon_l(\mathbf{k})}{\hbar\partial \mathbf{k}}\\
\label{diagonal}
\end{split}
\end{equation}
\begin{equation}
\begin{split}
H_{T,l\ne m}
&=\frac{i}{2\hbar}\mathbf{A}_{T}\cdot\boldsymbol{\mathcal{A}}_{lm}\left[\varepsilon_m^2(\mathbf{k})-\varepsilon_l^2(\mathbf{k})\right]\\
\label{nondiagonal}
\end{split}
\end{equation}
Then we substitute the Eq. (\ref{diagonal}) and (\ref{nondiagonal}) into the Eq. (\ref{aEQ})
\begin{equation}
\begin{split}
i\hbar\dot{a}_l(\mathbf{k},t)
&=-a_l(\mathbf{k},t)\varepsilon_l(\mathbf{k})\mathbf{A}_T\cdot\frac{\partial\varepsilon_l(\mathbf{k})}{\hbar\partial\mathbf{k}}+\frac{i}{2\hbar}\mathbf{A}_{T}\cdot\sum_{m\ne l}\boldsymbol{\mathcal{A}}_{lm}\left[\varepsilon_m^2(\mathbf{k})-\varepsilon_l^2(\mathbf{k})\right]a_m(\mathbf{k},t)e^{-i\left[\varepsilon_m(\mathbf{k})-\varepsilon_l(\mathbf{k})\right]t/\hbar}.\\
\end{split}
\end{equation}
In the main text, we take $\mathbf{A}_{T}=-\frac{\mathbf{E}_{T}}{\omega}\sin{\omega t}$ , so we can get
\begin{equation}
\begin{split}
\dot{a}_l(\mathbf{k},t)
&=-\frac{i\sin{\omega t}}{\hbar^2\omega}a_l(\mathbf{k},t)\varepsilon_l(\mathbf{k})\mathbf{E}_{T}\cdot\frac{\partial\varepsilon_l(\mathbf{k})}{\partial \mathbf{k}}
-\frac{\sin{\omega t}}{2\hbar^2\omega}\mathbf{E}_{T}\cdot\sum_{m\ne l}\boldsymbol{\mathcal{A}}_{lm}\left[\varepsilon_m^2(\mathbf{k})-\varepsilon_l^2(\mathbf{k})\right]a_m(\mathbf{k},t)e^{-i\varepsilon_{ml}(\mathbf{k})t/\hbar}.\\
\label{A9}
\end{split}
\end{equation}

We denote the zero-order perturbed part of $a_m(\mathbf{k},t)$ as $a_m^{(0)}(\mathbf{k},t)$, and denote the first-order perturbed part of $a_m(\mathbf{k},t)$ as $a_m^{(1)}(\mathbf{k},t)$. The initial condition is $a_m(\mathbf{k},0)=\delta_{mn}$, and the zeroth-order coefficients satisfy $\dot{a}_m^{(0)}(\mathbf{k},t)=0$. So we can get $a_m^{(0)}(\mathbf{k},t)=\delta_{mn}$. Substituting the zeroth-order coefficients into Eq. (\ref{A9}), we can know that the first-order coefficients satisfy
\begin{equation}
\begin{split}
\dot{a}_n^{(1)}(\mathbf{k},t)
&=-\frac{i\sin{\omega t}}{\hbar^2\omega}\varepsilon_n(\mathbf{k})\mathbf{E}_{T}\cdot\frac{\partial\varepsilon_n(\mathbf{k})}{\partial \mathbf{k}}
\end{split}
\end{equation}
\begin{equation}
\begin{split}
\dot{a}_{l\ne n}^{(1)}(\mathbf{k},t)
&=-\frac{\sin{\omega t}}{2\hbar^2\omega}\mathbf{E}_{T}\cdot\boldsymbol{\mathcal{A}}_{ln}\left[\varepsilon_n^2(\mathbf{k})-\varepsilon_l^2(\mathbf{k})\right]e^{-i\varepsilon_{nl}(\mathbf{k})t/\hbar}.\\
\end{split}
\end{equation}
Because $a_m(\mathbf{k},0)=\delta_{mn}$, the initial condition of the first-order coefficients satisfy $a_m^{(1)}(\mathbf{k},0)=0$. Then we can get the first-order coefficients
\begin{equation}
\begin{split}
a_n^{(1)}(\mathbf{k},t)=\int_0^tdt^{\prime}\dot{a}_n(\mathbf{k},t^{\prime})
=-i\frac{1}{\left(\hbar\omega\right)^2}\left(1-\cos{\omega t}\right)\varepsilon_n(\mathbf{k})\mathbf{E}_{T}\cdot\frac{\partial\varepsilon_n(\mathbf{k})}{\partial \mathbf{k}}
\end{split}
\end{equation}
\begin{equation}
\begin{split}
a_{l\ne n}^{(1)}(\mathbf{k},t)&=\int_0^tdt^{\prime}\dot{a}_{l\ne n}(\mathbf{k},t^{\prime})\\
&=\frac{\mathbf{E}_{T}}{4\hbar^2\omega}\cdot\boldsymbol{\mathcal{A}}_{ln}\left[\varepsilon_n^2(\mathbf{k})-\varepsilon_l^2(\mathbf{k})\right]\left[\frac{e^{i\left[\omega-\varepsilon_{nl}(\mathbf{k})/\hbar\right]t}-1}{\omega-\varepsilon_{nl}(\mathbf{k})/\hbar}+\frac{e^{-i\left[\omega+\varepsilon_{nl}(\mathbf{k})/\hbar\right]t}-1}{\omega+\varepsilon_{nl}(\mathbf{k})\hbar}\right].\\
\end{split}
\end{equation}
So the wave-function coefficients are
\begin{equation}
\begin{split}
a_n(\mathbf{k},t)=a_n^{(0)}(\mathbf{k},t)+a_n^{(1)}(\mathbf{k},t)
&=1-i\frac{1}{\left(\hbar\omega\right)^2}\left(1-\cos{\omega t}\right)\varepsilon_n(\mathbf{k})\mathbf{E}_{T}\cdot\frac{\partial\varepsilon_n(\mathbf{k})}{\partial \mathbf{k}}\approx e^{i\mathbf{E}_{T}\cdot\boldsymbol{\xi}_{n}(\mathbf{k})}
\label{an}
\end{split}
\end{equation}
\begin{equation}
\begin{split}
a_{l\ne n}(\mathbf{k},t)=a_{l\ne n}^{(0)}(\mathbf{k},t)+a_{l\ne n}^{(1)}(\mathbf{k},t)
&=\frac{\mathbf{E}_{T}}{4\hbar^2\omega}\cdot\boldsymbol{\mathcal{A}}_{ln}\left[\varepsilon_n^2(\mathbf{k})-\varepsilon_l^2(\mathbf{k})\right]\left[\frac{e^{i\left[\omega-\varepsilon_{nl}(\mathbf{k})/\hbar\right]t}-1}{\omega-\varepsilon_{nl}(\mathbf{k})/\hbar}+\frac{e^{-i\left[\omega+\varepsilon_{nl}(\mathbf{k})/\hbar\right] t}-1}{\omega+\varepsilon_{nl}(\mathbf{k})/\hbar}\right]\\
\label{al}
\end{split}
\end{equation}
in which $\boldsymbol{\xi}_{n}(\mathbf{k})=-\frac{1}{\left(\hbar\omega\right)^2}\left(1-\cos{\omega t}\right)\varepsilon_n(\mathbf{k})\frac{\partial\varepsilon_n(\mathbf{k})}{\partial \mathbf{k}}$.
Substituting (\ref{an}) and (\ref{al}) into the Eq. (\ref{Acorrbloch}), we can get the corrected Bloch state
\begin{equation}
\begin{split}
\ket{\tilde{u}_n(\mathbf{k})}
&=e^{-i\varepsilon_n(\mathbf{k})t/\hbar}e^{i\mathbf{E}_{T}\cdot\boldsymbol{\xi}_{n}(\mathbf{k})}\ket{u_n(\mathbf{k})}\\
&+\sum_{l\ne n}\left\{\frac{\mathbf{E}_{T}}{4\hbar^2\omega}\cdot\boldsymbol{\mathcal{A}}_{ln}\left[\varepsilon_n^2(\mathbf{k})-\varepsilon_l^2(\mathbf{k})\right]\left[\frac{e^{i\left[\omega-\varepsilon_{nl}(\mathbf{k})/\hbar\right]t}-1}{\omega-\varepsilon_{nl}(\mathbf{k})/\hbar}+\frac{e^{-i\left[\omega+\varepsilon_{nl}(\mathbf{k})/\hbar\right] t}-1}{\omega+\varepsilon_{nl}(\mathbf{k})/\hbar}\right]\right\}e^{-i\varepsilon_l(\mathbf{k})t/\hbar}\ket{u_l(\mathbf{k})}.\\
\end{split}
\end{equation}
For simplicity, The phase transformation $e^{i\left[\varepsilon_n(\mathbf{k})t/\hbar-\mathbf{E}_{T}\cdot\mathbf{\xi}_{n}(\mathbf{k})\right]}$ is taken on the corrected Bloch state, and the result is accurate to the first order of the thermal field $\mathbf{E}_{T}$
\begin{equation}
\begin{split}
\ket{\tilde{u}_n(\mathbf{k})}&\to e^{i\left[\varepsilon_n(\mathbf{k})t/\hbar-\mathbf{E}_{T}\cdot\mathbf{\xi}_{n}(\mathbf{k})\right]}\ket{\tilde{u}_n(\mathbf{k})}\\
&=\ket{u_n(\mathbf{k})}\\
&+\sum_{l\ne n}\left\{\frac{\mathbf{E}_{T}}{4\hbar^2\omega}\cdot\boldsymbol{\mathcal{A}}_{ln}\left[\varepsilon_n^2(\mathbf{k})-\varepsilon_l^2(\mathbf{k})\right]\left[\frac{e^{i\left[\omega-\varepsilon_{nl}(\mathbf{k})/\hbar\right]t}-1}{\omega-\varepsilon_{nl}(\mathbf{k})/\hbar}+\frac{e^{-i\left[\omega+\varepsilon_{nl}(\mathbf{k})/\hbar\right] t}-1}{\omega+\varepsilon_{nl}(\mathbf{k})/\hbar}\right]\right\}e^{-i\varepsilon_{ln}(\mathbf{k})t/\hbar}e^{-i\mathbf{E}_{T}\cdot\boldsymbol{\xi}_{n}(\mathbf{k})}\ket{u_l(\mathbf{k})}\\
&\approx\ket{u_n(\mathbf{k})}\\
&+\sum_{l\ne n}\left\{\frac{\mathbf{E}_{T}}{4\hbar^2\omega}\cdot\boldsymbol{\mathcal{A}}_{ln}\left[\varepsilon_n^2(\mathbf{k})-\varepsilon_l^2(\mathbf{k})\right]\left[\frac{e^{i\left[\omega-\varepsilon_{nl}(\mathbf{k})/\hbar\right]t}-1}{\omega-\varepsilon_{nl}(\mathbf{k})/\hbar}+\frac{e^{-i\left[\omega+\varepsilon_{nl}(\mathbf{k})/\hbar\right] t}-1}{\omega+\varepsilon_{nl}(\mathbf{k})/\hbar}\right]\right\}\\
&\cdot e^{-i\varepsilon_{ln}(\mathbf{k})t/\hbar}\left(1-i\mathbf{E}_{T}\cdot\boldsymbol{\xi}_{n}(\mathbf{k})\right)\ket{u_l(\mathbf{k})}.\\
\label{ptcorrbloch}
\end{split}
\end{equation}
We only take the first-order term of the thermal field $\mathbf{E}_T$ in Eq. (\ref{ptcorrbloch})
\begin{equation}
\begin{split}
\ket{\tilde{u}_n(\mathbf{k})}
&\approx\ket{u_n(\mathbf{k})}\\
&+\sum_{l\ne n}\left\{\frac{\mathbf{E}_{T}}{4\hbar^2\omega}\cdot\boldsymbol{\mathcal{A}}_{ln}\left[\varepsilon_n^2(\mathbf{k})-\varepsilon_l^2(\mathbf{k})\right]\left[\frac{e^{i\left[\omega-\varepsilon_{nl}(\mathbf{k})/\hbar\right]t}-1}{\omega-\varepsilon_{nl}(\mathbf{k})/\hbar}+\frac{e^{-i\left[\omega+\varepsilon_{nl}(\mathbf{k})/\hbar\right] t}-1}{\omega+\varepsilon_{nl}(\mathbf{k})/\hbar}\right]\right\}e^{-i\varepsilon_{ln}(\mathbf{k})t/\hbar}\ket{u_l(\mathbf{k})}.\\
\end{split}
\end{equation}

Then we take the zero frequency limit $\omega\to 0$, and remove non-physical items that increase over time \cite{ref18}
\begin{equation}
\begin{split}
\ket{\tilde{u}_n(\mathbf{k})}
=\ket{u_n(\mathbf{k})}-\sum_{l\ne n}\frac{\mathbf{E}_{T}}{2}\cdot\boldsymbol{\mathcal{A}}_{ln}\left(1-e^{i\varepsilon_{nl}(\mathbf{k})t/\hbar}\right)\frac{\varepsilon_n(\mathbf{k})+\varepsilon_l(\mathbf{k})}{\varepsilon_n(\mathbf{k})-\varepsilon_l(\mathbf{k})}\ket{u_l(\mathbf{k})}.
\end{split}
\end{equation}
To get a steady-state, we should introduce a relaxation $\tau$ due to scattering. We take the long-time approximation $(t\gg\tau)$, such that all oscillatory terms drop out \cite{ref18}. Therefore, we can get the corrected Bloch state
\begin{equation}
\ket{\tilde{u}_n(\mathbf{k})}
=\ket{u_n(\mathbf{k})}-\sum_{l\ne n}\frac{\mathbf{E}_{T}}{2}\cdot\boldsymbol{\mathcal{A}}_{ln}\frac{\varepsilon_n(\mathbf{k})+\varepsilon_l(\mathbf{k})}{\varepsilon_n(\mathbf{k})-\varepsilon_l(\mathbf{k})}\ket{u_l(\mathbf{k})}.
\label{cb}
\end{equation}

\section{Corrected Bloch state derived by the TSP method} \label{AB}

We derive the perturbed Bloch state using TSP method. Our result is also accurate to the first order of temperature gradient.

\subsubsection{The perturbation from temperature gradient}
We describe temperature gradient by thermal scalar potential
\begin{equation}
\mathbf{E}_T=-\boldsymbol{\nabla}\Psi(\mathbf{r})=-\frac{\boldsymbol{\nabla} T}{T}\approx-\frac{\boldsymbol{\nabla} T}{T_0}.
\end{equation}

Analogy the gravitational potential in electron system \cite{Smrcka1977},  Hamiltonian under the temperature gradient can be described as
\begin{equation}
\begin{split}
\hat{H}_{T}&=\hat{H}_0+\frac{1}{2}\left[\hat{H}_0,\hat{\mathbf{r}}\right]_+\cdot\boldsymbol{\nabla}\Psi(\mathbf{r})\\
&=\hat{H}_0-\frac{1}{2}\left[\hat{H}_0,\hat{\mathbf{r}}\right]_+\cdot\mathbf{E}_{T},\\
\end{split}
\end{equation}
 in which $\hat{H}_0$ is the Hamiltonian that describe magnon system without temperature gradient. Here, we take $c=1$.
Then we take
\begin{equation}
\begin{split}
\hat{H}_T(\mathbf{k})&=e^{-i\mathbf{k}\cdot\mathbf{r}}\hat{H}_Te^{i\mathbf{k}\cdot\mathbf{r}}\\
&=\hat{H}_0(\mathbf{k})-\frac{1}{2}\left[\hat{H}_0(\mathbf{k}),\hat{\mathbf{r}}\right]_+\cdot\mathbf{E}_{T},\\
\end{split}
\end{equation}
in which $\hat{H}^{\prime}(\mathbf{k})=-\frac{1}{2}\left[\hat{H}_0(\mathbf{k}),\hat{\mathbf{r}}\right]_+\cdot\mathbf{E}_{T}$ is the perturbation. And $\hat{H}_0(\mathbf{k})$ satisfies
\begin{equation}
\hat{H}_0(\mathbf{k})\ket{u_n(\mathbf{k})}=\varepsilon_n(\mathbf{k})\ket{u_n(\mathbf{k})}
\end{equation}
in which $\ket{u_n(\mathbf{k})}$ is Bloch state of magnons.

\subsubsection{The corrected Bloch state}

We follow the method in Ref. \onlinecite{ref5}, and derive the modified Bloch state. Without the perturbation, the wave packet can be expressed as
\begin{equation}
\ket{W_n}=\int[dk]C_n(\mathbf{k},t)e^{i\mathbf{k}\cdot\mathbf{r}}\ket{u_n(\mathbf{k})}.
\end{equation}
Considering the influence of the perturbation term $\hat{H}^{\prime}(\mathbf{k})$, the corrected Bloch state is
\begin{equation}
\begin{split}
\ket{\tilde{u}_n(\mathbf{k})}&=\ket{u_n(\mathbf{k})}+\ket{u_n^{\prime}(\mathbf{k})}\\
&=\ket{u_n(\mathbf{k})}+\sum_{m\ne n}a_m(\mathbf{k})\ket{u_m(\mathbf{k})}.\\
\end{split}
\end{equation}
Then the state of corrected wave packet turns to
\begin{equation}
\begin{split}
\ket{W_n}&=\int[dk]C_n(\mathbf{k},t)e^{i\mathbf{k}\cdot\mathbf{r}}\ket{\tilde{u}_n(\mathbf{k})}\\
&=\int[dk]C_n(\mathbf{k},t)e^{i\mathbf{k}\cdot\mathbf{r}}\left[\ket{u_n(\mathbf{k})}+\ket{u_n^{\prime}(\mathbf{k})}\right]\\
&=\int[dk]C_n(\mathbf{k},t)e^{i\mathbf{k}\cdot\mathbf{r}}\left[\ket{u_n(\mathbf{k})}+\sum_{m\ne n}a_m(\mathbf{k})\ket{u_m(\mathbf{k})}\right]\\
&=\int[dk]\left[C_n(\mathbf{k},t)e^{i\mathbf{k}\cdot\mathbf{r}}\ket{u_n(\mathbf{k})}+\sum_{m\ne n}C_m(\mathbf{k},t)e^{i\mathbf{k}\cdot\mathbf{r}}\ket{u_m(\mathbf{k})}\right]\\
\label{wavepacket}
\end{split}
\end{equation}
in which $C_{m\ne n}(\mathbf{k},t)=C_n(\mathbf{k},t)a_m(\mathbf{k})$.

To calculate the wave packet coefficients $C_m(\mathbf{k},t)$, we plug Eq. (\ref{wavepacket}) into time-dependent Schr\"{o}dinger equation
\begin{equation}
i\hbar\partial_t\ket{W_n}=\left(\hat{H}_0+\hat{H}^{\prime}\right)\ket{W_n}.
\label{TDSQTSP}
\end{equation}

We start with the left-hand side of Eq. (\ref{TDSQTSP})
\begin{equation}
\begin{split}
i\hbar\partial_t\ket{W_n}
&=i\hbar\int[dk]e^{i\mathbf{k}\cdot\mathbf{r}}\left[\dot{C}_n(\mathbf{k},t)\ket{u_n(\mathbf{k})}+\sum_{m\ne n}\dot{C}_n(\mathbf{k},t)a_m(\mathbf{k})\ket{u_m(\mathbf{k})}\right]\\
&=\int[dk]e^{i\mathbf{k}\cdot\mathbf{r}}\left[\epsilon_n(\mathbf{k})C_n(\mathbf{k},t)\ket{u_n(\mathbf{k})}+\sum_{m\ne n}\epsilon_n(\mathbf{k})C_n(\mathbf{k},t)a_m(\mathbf{k})\ket{u_m(\mathbf{k})}\right]\\
&=\int[dk]\varepsilon_n(\mathbf{k})\left[C_n(\mathbf{k},t)e^{i\mathbf{k}\cdot\mathbf{r}}\ket{u_n(\mathbf{k})}+\sum_{m\ne n}C_m(\mathbf{k},t)e^{i\mathbf{k}\cdot\mathbf{r}}\ket{u_m(\mathbf{k})}\right].\\
\end{split}
\end{equation}
Then we take the inner product $\bra{u_l(\mathbf{k}^{\prime})}e^{-i\mathbf{k}^{\prime}\cdot\mathbf{r}}$ on the above equation
\begin{equation}
\begin{aligned}
\bra{u_l(\mathbf{k}^{\prime})}e^{-i\mathbf{k}^{\prime}\cdot\mathbf{r}}i\hbar\partial_t\ket{W_n}
&=\int[dk]\varepsilon_n(\mathbf{k})\left[C_n(\mathbf{k},t)\bra{u_l(\mathbf{k}^{\prime})}e^{-i\mathbf{k}^{\prime}\cdot\mathbf{r}}e^{i\mathbf{k}\cdot\mathbf{r}}\ket{u_n(\mathbf{k})} \right.\\
&+\left.\sum_{m\ne n}C_m(\mathbf{k},t)\bra{u_l(\mathbf{k}^{\prime})}e^{-i\mathbf{k}^{\prime}\cdot\mathbf{r}}e^{i\mathbf{k}\cdot\mathbf{r}}\ket{u_m(\mathbf{k})}\right]\\
&=\varepsilon_n(\mathbf{k}^{\prime})C_l(\mathbf{k}^{\prime},t),
\end{aligned}
\end{equation}
in which $l\ne n$.
Now we derive the right-hand side of Eq. (\ref{TDSQTSP})
\begin{equation}
\begin{split}
\left(\hat{H}_0+\hat{H}^{\prime}\right)\ket{W_n}&=\int[dk]e^{i\mathbf{k}\cdot\mathbf{r}}\left\{C_n(\mathbf{k},t)\left[\hat{H}_0(\mathbf{k})
+\hat{H}^{\prime}(\mathbf{k})\right]\ket{u_n(\mathbf{k})}+\sum_{m\ne n}C_m(\mathbf{k},t)\left[\hat{H}_0(\mathbf{k})+\hat{H}^{\prime}(\mathbf{k})\right]\ket{u_m(\mathbf{k})}\right\}\\
&=\int[dk]e^{i\mathbf{k}\cdot\mathbf{r}}\left[C_n(\mathbf{k},t)\varepsilon_n(\mathbf{k})\ket{u_n(\mathbf{k})}+\sum_{m\ne n}C_m(\mathbf{k},t)\varepsilon_m(\mathbf{k})\ket{u_m(\mathbf{k})}\right]\\
&+\int[dk]e^{i\mathbf{k}\cdot\mathbf{r}}\left[C_n(\mathbf{k},t)\hat{H}^{\prime}(\mathbf{k})\ket{u_n(\mathbf{k})}+\sum_{m\ne n}C_m(\mathbf{k},t)\hat{H}^{\prime}(\mathbf{k})\ket{u_m(\mathbf{k})}\right]\\
\end{split}
\end{equation}
Then we take the inner product $\bra{u_l(\mathbf{k}^{\prime})}e^{-i\mathbf{k}^{\prime}\cdot\mathbf{r}}$ on the above equation ($l\ne n$)
\begin{equation}
\begin{split}
&\bra{u_l(\mathbf{k}^{\prime})}e^{-i\mathbf{k}^{\prime}\cdot\mathbf{r}}\left(\hat{H}_0+\hat{H}^{\prime}\right)\ket{W_n}\\
&=\int[dk]\left[C_n(\mathbf{k},t)\varepsilon_n(\mathbf{k})\bra{u_l(\mathbf{k}^{\prime})}e^{i(\mathbf{k}-\mathbf{k}^{\prime})\cdot\mathbf{r}}\ket{u_n(\mathbf{k})}+\sum_{m\ne n}C_m(\mathbf{k},t)\varepsilon_m(\mathbf{k})\bra{u_l(\mathbf{k}^{\prime})}e^{i(\mathbf{k}-\mathbf{k}^{\prime})\cdot\mathbf{r}}\ket{u_m(\mathbf{k})}\right]\\
&+\int[dk]\left[C_n(\mathbf{k},t)\bra{u_l(\mathbf{k}^{\prime})}e^{i(\mathbf{k}-\mathbf{k}^{\prime})\cdot\mathbf{r}}\hat{H}^{\prime}(\mathbf{k})\ket{u_n(\mathbf{k})}+\sum_{m\ne n}C_m(\mathbf{k},t)\bra{u_l(\mathbf{k}^{\prime})}e^{i(\mathbf{k}-\mathbf{k}^{\prime})\cdot\mathbf{r}}\hat{H}^{\prime}(\mathbf{k})\ket{u_m(\mathbf{k})}\right]\\
&=C_l(\mathbf{k}^{\prime},t)\varepsilon_l(\mathbf{k}^{\prime})\\
&-\frac{1}{2}\mathbf{E}_{T}\cdot\int[dk]C_n(\mathbf{k},t)\bra{u_l(\mathbf{k}^{\prime})}e^{i(\mathbf{k}-\mathbf{k}^{\prime})\cdot\mathbf{r}}\left[\hat{H}_0(\mathbf{k}),\hat{\mathbf{r}}\right]_+\ket{u_n(\mathbf{k})}\\
&-\frac{1}{2}\mathbf{E}_{T}\cdot\sum_{m\ne n}\int[dk]C_m(\mathbf{k},t)\bra{u_l(\mathbf{k}^{\prime})}e^{i(\mathbf{k}-\mathbf{k}^{\prime})\cdot\mathbf{r}}\left[\hat{H}_0(\mathbf{k}),\hat{\mathbf{r}}\right]_+\ket{u_m(\mathbf{k})}\\
&=C_l(\mathbf{k}^{\prime},t)\varepsilon_l(\mathbf{k}^{\prime})\\
&-\frac{1}{2}\mathbf{E}_{T}\cdot\int[dk]C_n(\mathbf{k},t)\bra{u_l(\mathbf{k}^{\prime})}\hat{H}_0(\mathbf{k}^{\prime})\hat{\mathbf{r}}e^{i(\mathbf{k}-\mathbf{k}^{\prime})\cdot\mathbf{r}}\ket{u_n(\mathbf{k})}\\
&-\frac{1}{2}\mathbf{E}_{T}\cdot\int[dk]C_n(\mathbf{k},t)\bra{u_l(\mathbf{k}^{\prime})}e^{i(\mathbf{k}-\mathbf{k}^{\prime})\cdot\mathbf{r}}\hat{\mathbf{r}}\hat{H}_0(\mathbf{k})\ket{u_n(\mathbf{k})}\\
&-\frac{1}{2}\mathbf{E}_{T}\cdot\sum_{m\ne n}\int[dk]C_m(\mathbf{k},t)\bra{u_l(\mathbf{k}^{\prime})}\hat{H}_0(\mathbf{k}^{\prime})\hat{\mathbf{r}}e^{i(\mathbf{k}-\mathbf{k}^{\prime})\cdot\mathbf{r}}\ket{u_m(\mathbf{k})}\\
&-\frac{1}{2}\mathbf{E}_{T}\cdot\sum_{m\ne n}\int[dk]C_m(\mathbf{k},t)\bra{u_l(\mathbf{k}^{\prime})}e^{i(\mathbf{k}-\mathbf{k}^{\prime})\cdot\mathbf{r}}\hat{\mathbf{r}}\hat{H}_0(\mathbf{k})\ket{u_m(\mathbf{k})}\\
&=C_l(\mathbf{k}^{\prime},t)\varepsilon_l(\mathbf{k}^{\prime})\\
&-\frac{1}{2}\mathbf{E}_{T}\cdot\int[dk]C_n(\mathbf{k},t)\left[\varepsilon_l(\mathbf{k}^{\prime})+\varepsilon_n(\mathbf{k})\right]\bra{u_l(\mathbf{k}^{\prime})}\hat{\mathbf{r}}e^{i(\mathbf{k}-\mathbf{k}^{\prime})\cdot\mathbf{r}}\ket{u_n(\mathbf{k})}\\
&-\frac{1}{2}\mathbf{E}_{T}\cdot\sum_{m\ne n}\int[dk]C_m(\mathbf{k},t)\left[\varepsilon_l(\mathbf{k}^{\prime})+\varepsilon_m(\mathbf{k})\right]\bra{u_l(\mathbf{k}^{\prime})}\hat{\mathbf{r}}e^{i(\mathbf{k}-\mathbf{k}^{\prime})\cdot\mathbf{r}}\ket{u_m(\mathbf{k})}\\
&=C_l(\mathbf{k}^{\prime},t)\varepsilon_l(\mathbf{k}^{\prime})-\frac{1}{2}\mathbf{E}_{T}\cdot\sum_{m}\int[dk]C_m(\mathbf{k},t)\left[\varepsilon_l(\mathbf{k}^{\prime})+\varepsilon_m(\mathbf{k})\right]\bra{\psi_l(\mathbf{k}^{\prime})}\hat{\mathbf{r}}\ket{\psi_m(\mathbf{k})}. \\
\end{split}
\end{equation}
From \cite{ref2}, we know $\bra{\psi_l(\mathbf{k}^{\prime})}\hat{\mathbf{r}}\ket{\psi_m(\mathbf{k})}=i\partial_{\mathbf{k}^{\prime}}\delta(\mathbf{k}-\mathbf{k}^{\prime})\delta_{lm}+\boldsymbol{\mathcal{A}}_{lm}\delta(\mathbf{k}-\mathbf{k}^{\prime})$, so we can get
\begin{equation}
\begin{aligned}
&\bra{u_l(\mathbf{k}^{\prime})}e^{-i\mathbf{k}^{\prime}\cdot\mathbf{r}}\left(\hat{H}_0+\hat{H}^{\prime}\right)\ket{W_n}\\
&=\int[dk]C_l(\mathbf{k}^{\prime},t)\varepsilon_l(\mathbf{k}^{\prime})-\frac{1}{2}\mathbf{E}_{T}\cdot\sum_{m}\int[dk]C_m(\mathbf{k},t)\left[\varepsilon_l(\mathbf{k}^{\prime})
+\varepsilon_m(\mathbf{k})\right]\left[i\partial_{\mathbf{k}^{\prime}}\delta(\mathbf{k}-\mathbf{k}^{\prime})\delta_{lm}+\boldsymbol{\mathcal{A}}_{lm}\delta(\mathbf{k}-\mathbf{k}^{\prime})\right]\\
&=\int[dk]C_l(\mathbf{k}^{\prime},t)\varepsilon_l(\mathbf{k}^{\prime})-\frac{1}{2}\mathbf{E}_{T}\cdot\sum_{m}\varepsilon_l(\mathbf{k}^{\prime})i\partial_{\mathbf{k}^{\prime}}
\int[dk]C_m(\mathbf{k},t)\delta(\mathbf{k}-\mathbf{k}^{\prime})\delta_{lm} \\
&-\frac{1}{2}\mathbf{E}_{T}\cdot\sum_{m}i\partial_{\mathbf{k}^{\prime}}\int[dk]C_m(\mathbf{k})\varepsilon_m(\mathbf{k})\delta(\mathbf{k}-\mathbf{k}^{\prime})\delta_{lm}
 -\frac{1}{2}\mathbf{E}_{T}\cdot\sum_{m}\int[dk]\boldsymbol{\mathcal{A}}_{lm}\delta(\mathbf{k}-\mathbf{k}^{\prime})C_m(\mathbf{k},t)\left[\varepsilon_l(\mathbf{k}^{\prime})+\varepsilon_m(\mathbf{k})\right]\\
&=\int[dk]C_l(\mathbf{k}^{\prime},t)\varepsilon_l(\mathbf{k}^{\prime})
-\frac{1}{2}\mathbf{E}_{T}\cdot\varepsilon_l(\mathbf{k}^{\prime})i\partial_{\mathbf{k}^{\prime}}C_l(\mathbf{k}^{\prime},t)
-\frac{1}{2}\mathbf{E}_{T}\cdot i\partial_{\mathbf{k}^{\prime}}\left[C_l(\mathbf{k}^{\prime},t)\varepsilon_l(\mathbf{k}^{\prime})\right] \\
&-\frac{1}{2}\mathbf{E}_{T}\cdot\sum_{m}\boldsymbol{\mathcal{A}}_{lm}C_m(\mathbf{k}^{\prime},t)\left[\varepsilon_l(\mathbf{k}^{\prime})+\varepsilon_m(\mathbf{k}^{\prime})\right].\\
\label{righthand}
\end{aligned}
\end{equation}
Here, $\delta(\mathbf{k}-\mathbf{k}^{\prime})$ satisfies $\int[dk]\delta(\mathbf{k}-\mathbf{k}^{\prime})=1$. For the band $l$ ($l\ne n$), the wave-function coefficient can be expressed as $C_{l\ne n}(\mathbf{k},t)=C_n(\mathbf{k},t)a_l(\mathbf{k})$, in which $a_l(\mathbf{k})$ is from the perturbation of  temperature gradient. Therefore, $C_{l\ne n}(\mathbf{k},t)$ contains the thermal field $\mathbf{E}_{T}$. Because our result is only accurate to the first order of
temperature gradient, we get rid of the higher order terms of $\mathbf{E}_{T}$ in Eq. (\ref{righthand}).  Then we can get
\begin{equation}
\bra{u_l(\mathbf{k}^{\prime})}e^{-i\mathbf{k}^{\prime}\cdot\mathbf{r}}\left(\hat{H}_0+\hat{H}^{\prime}\right)\ket{W_n}=\int[dk]C_l(\mathbf{k}^{\prime},t)\varepsilon_l(\mathbf{k}^{\prime}) -\frac{1}{2}\mathbf{E}_{T}\cdot\boldsymbol{\mathcal{A}}_{ln} C_n(\mathbf{k}^{\prime},t)\left[\varepsilon_l(\mathbf{k}^{\prime})+\varepsilon_n(\mathbf{k}^{\prime})\right].
\end{equation}
From time-dependent Schr\"{o}dinger equation (Eq. (\ref{TDSQTSP})), we can get
\begin{equation}
\varepsilon_n(\mathbf{k})C_l(\mathbf{k},t)=\varepsilon_l(\mathbf{k})C_l(\mathbf{k},t) -\frac{1}{2}\mathbf{E}_{T}\cdot\boldsymbol{\mathcal{A}}_{ln} C_n(\mathbf{k},t)\left[\varepsilon_l(\mathbf{k})+\varepsilon_n(\mathbf{k})\right].
\end{equation}
Then we can get the wave packet coefficients
\begin{equation}
C_l(\mathbf{k},t)=C_n(\mathbf{k},t)a_l(\mathbf{k})=-\frac{1}{2}\mathbf{E}_{T}\cdot\boldsymbol{\mathcal{A}}_{ln} C_n(\mathbf{k},t)\frac{\varepsilon_n(\mathbf{k})+\varepsilon_l(\mathbf{k})}{\varepsilon_n(\mathbf{k})-\varepsilon_l(\mathbf{k})},
\end{equation}
that is to say
\begin{equation}
a_l(\mathbf{k})=-\frac{1}{2}\mathbf{E}_{T}\cdot\boldsymbol{\mathcal{A}}_{ln}\frac{\varepsilon_n(\mathbf{k})+\varepsilon_l(\mathbf{k})}{\varepsilon_n(\mathbf{k})-\varepsilon_l(\mathbf{k})},
\end{equation}
and the corrected Bloch state is
\begin{equation}
\ket{\tilde{u}_n(\mathbf{k})}
=\ket{u_n(\mathbf{k})}-\sum_{l\ne n}\frac{\mathbf{E}_{T}}{2}\cdot\boldsymbol{\mathcal{A}}_{ln}\frac{\varepsilon_n(\mathbf{k})+\varepsilon_l(\mathbf{k})}{\varepsilon_n(\mathbf{k})-\varepsilon_l(\mathbf{k})}\ket{u_l(\mathbf{k})}.
\end{equation}
which is same as the result from the TVP method (Eq. \ref{corrbloch}).

\section{The correction of Berry connection and the thermal Berry connection polarizability} \label{AC}

Now we derive the specific expression for the correction of Berry connection
\begin{equation}
\begin{split}
\mathcal{A}_{n,j}^{\prime}(\mathbf{k})
&=\bra{u_n(\mathbf{k})}i\partial_{k_j}\ket{u_n^{\prime}(\mathbf{k})}+\bra{u_n^{\prime}(\mathbf{k})}i\partial_{k_j}\ket{u_n(\mathbf{k})}\\
&=-\frac{1}{2}\sum_{l\ne n}\bra{u_n(\mathbf{k})}i\partial_{k_j}\mathcal{A}_{ln,i}\frac{\varepsilon_n(\mathbf{k})+\varepsilon_l(\mathbf{k})}{\varepsilon_n(\mathbf{k})-\varepsilon_l(\mathbf{k})}\ket{u_l(\mathbf{k})}E_{T,i}+h.c.\\
&=-\frac{1}{2}\sum_{l\ne n}\left[i\partial_{k_j}\mathcal{A}_{ln,i}\frac{\varepsilon_n(\mathbf{k})+\varepsilon_l(\mathbf{k})}{\varepsilon_n(\mathbf{k})-\varepsilon_l(\mathbf{k})}\right]\delta_{nl}E_{T,i}-\frac{1}{2}\sum_{l\ne n}\mathcal{A}_{ln,i}\frac{\varepsilon_n(\mathbf{k})+\varepsilon_l(\mathbf{k})}{\varepsilon_n(\mathbf{k})-\varepsilon_l(\mathbf{k})}\bra{u_n(\mathbf{k})}i\partial_{k_j}\ket{u_l(\mathbf{k})}E_{T,i}+h.c.\\
&=-\frac{1}{2}\sum_{l\ne n}\mathcal{A}_{ln,i}\frac{\varepsilon_n(\mathbf{k})+\varepsilon_l(\mathbf{k})}{\varepsilon_n(\mathbf{k})-\varepsilon_l(\mathbf{k})}\bra{u_n(\mathbf{k})}i\partial_{k_j}\ket{u_l(\mathbf{k})}E_{T,i}+h.c.\\
&=-\frac{1}{2}\sum_{l\ne n}\left[\varepsilon_n(\mathbf{k})+\varepsilon_l(\mathbf{k})\right]\frac{\mathcal{A}_{nl,j}\mathcal{A}_{ln,i}}{\varepsilon_n(\mathbf{k})-\varepsilon_l(\mathbf{k})}E_{T,i}+h.c.\\
&=-Re\left\{\sum_{l\ne n}\left[\varepsilon_n(\mathbf{k})+\varepsilon_l(\mathbf{k})\right]\frac{\mathcal{A}_{nl,j}\mathcal{A}_{ln,i}}{\varepsilon_n(\mathbf{k})-\varepsilon_l(\mathbf{k})}\right\}E_{T,i}\\
&=G_{n,ji}^t(\mathbf{k})E_{T,i}
\end{split}
\end{equation}
in which
\begin{equation}
\begin{split}
G_{n,ji}^t(\mathbf{k})=-Re\left\{\sum_{l\ne n}\left[\varepsilon_n(\mathbf{k})+\varepsilon_l(\mathbf{k})\right]\frac{\mathcal{A}_{nl,j}\mathcal{A}_{ln,i}}{\varepsilon_n(\mathbf{k})-\varepsilon_l(\mathbf{k})}\right\}
\end{split}
\end{equation}
is the thermal Berry connection polarizability (TBCP). The subscripts $i$ and $j$ means Cartesian coordinate (x,y and z). And we take Einstein summation convention in this letter.

\section{The derivation of the heat current density correction induced by TBCP} \label{HeatTBCP}

Following the Eq. (\ref{jcor}) in the main text, we can get
\begin{equation}
\begin{split}
j_{Qi}^{\text{corrected}}&=-\epsilon_{ijz}T\left[\partial_j(\frac{1}{T})\right]\frac{k_B^2T^2}{\hbar V}\sum_{n\mathbf{k}}\Omega_{n,z}^{\prime}(\mathbf{k})c_2(\rho_n^B)\\
&=-\epsilon_{ijz}T\left[\partial_j(\frac{1}{T})\right]\frac{k_B^2T^2}{\hbar V}\sum_{n\mathbf{k}}\left[\epsilon_{\alpha\beta z}\partial_{k_{\alpha}}G_{n,\beta\delta}^t(\mathbf{k})E_{T,\delta}\right]c_2(\rho_n^B)\\
&=-\epsilon_{ijz}\epsilon_{\alpha\beta z}T\left[\partial_j(\frac{1}{T})\right]E_{T,\delta}\frac{k_B^2T^2}{\hbar V}\sum_{n\mathbf{k}}c_2(\rho_n^B)\partial_{k_{\alpha}}G_{n,\beta\delta}^t(\mathbf{k})\\
&=-\epsilon_{ijz}\epsilon_{\alpha\beta z}\frac{(\partial_jT)(\partial_{\delta}T)}{TT_0}\frac{k_B^2T^2}{\hbar V}\sum_{n\mathbf{k}}c_2(\rho_n^B)\partial_{k_{\alpha}}G_{n,\beta\delta}^t(\mathbf{k})\\
&=(\delta_{i\beta}\delta_{j\alpha}-\delta_{i\alpha}\delta_{j\beta})\frac{(\partial_jT)(\partial_{\delta}T)}{TT_0}\frac{k_B^2T^2}{\hbar V}\sum_{n\mathbf{k}}c_2(\rho_n^B)\partial_{k_{\alpha}}G_{n,\beta\delta}^t(\mathbf{k})\\
&=\frac{(\partial_jT)(\partial_{\delta}T)}{T_0}\frac{k_B^2T}{\hbar V}\sum_{n\mathbf{k}}c_2(\rho_n^B)\left[\partial_{k_{j}}G_{n, i\delta}^t(\mathbf{k})-\partial_{k_{i}}G_{n, j\delta}^t(\mathbf{k})\right]\\
&\approx\frac{(\partial_jT)(\partial_{\delta}T)}{T_0}\frac{k_B^2\left(T_0+r_{\gamma}\partial_{\gamma}T\right)}{\hbar V}\sum_{n\mathbf{k}}c_2(\rho_n^B)\left[\partial_{k_{j}}G_{n, i\delta}^t(\mathbf{k})-\partial_{k_{i}}G_{n, j\delta}^t(\mathbf{k})\right].
\end{split}
\label{AppJQi}
\end{equation}
From Eq. (\ref{AppJQi}) and keeping to the second order of gradient of temperature, we get the intrinsic second-order Hall heat current of magnon as
\begin{equation}
\begin{split}
j_{Qi}^{(2)}
&=(\partial_jT)(\partial_{\delta}T)\frac{k_B^2}{\hbar V}\sum_{n\mathbf{k}}c_2(\rho_n^B)\left[\partial_{k_{j}}G_{n, i\delta}^t(\mathbf{k})-\partial_{k_{i}}G_{n,j\delta}^t(\mathbf{k})\right],\\
\end{split}
\end{equation}
then we can get the intrinsic second-order magnon thermal Hall conductivity as Eq. (30) in the main text. Here we take $T\approx T_0$ in the Bose distribution $\rho_n^B$.
\section{Further details of model} \label{AE}

In this section, we take quadratic quantization of the Hamiltonian.
\begin{equation}
\begin{split}
\hat{H}&=\sum_{\langle ij\rangle}J_{ij}\hat{\mathbf{S}}_i\cdot\hat{\mathbf{S}}_j+\sum_{\ll ij\gg}\mathbf{D}_{ij}\cdot\left(\hat{\mathbf{S}}_i\times\hat{\mathbf{S}}_j\right)+g_{J}\mu_B\sum_i\hat{\mathbf{S}}_i\cdot\mathbf{B} (\mathbf{r})+\mathcal{K}\sum_i\hat{S}_{i}^{z2}\\
\end{split}
\end{equation}
We divide the Hamiltonian into three parts, $\hat{H}_1=\sum_{\langle ij\rangle}J_{ij}\hat{\mathbf{S}}_i\cdot\hat{\mathbf{S}}_j$, $\hat{H}_2=\sum_{\ll ij\gg}\mathbf{D}_{ij}\cdot\left(\hat{\mathbf{S}}_i\times\hat{\mathbf{S}}_j\right)$, $\hat{H}_3=g_{J}\mu_B\sum_i\hat{\mathbf{S}}_i\cdot\mathbf{B}(\mathbf{r})$, and $\hat{H}_4=\mathcal{K}\sum_i\hat{S}_i^{z2}$.

In the Hexagonal lattice model, the real space lattice basis vectors are
\begin{equation}
\begin{split}
\mathbf{a}_1=\frac{a}{2}\left(\sqrt{3},1\right), \mathbf{a}_2=\frac{a}{2}\left(\sqrt{3},-1\right).
\end{split}
\end{equation}
Then, $a=1$ nm is taken.

For Ferromagnetic Hexagonal lattice, we need to apply a periodic magnetic field to break symmetry $\mathcal{P}$. Under the periodic magnetic field, the red and blue lattice points are subjected to opposite magnetic field. Firstly, we take the Holstein-Primakoff transform
\begin{equation}
\begin{cases}
&\hat{S}_{A,i}^{+}=\sqrt{2S}\hat{a}_i\\
&\hat{S}_{A,i}^{-}=\sqrt{2S}\hat{a}_i^{\dag}\\
&\hat{S}_{A,i}^{z}=S-\hat{a}_i^{\dag}\hat{a}_i\\
\end{cases},
\\
\end{equation}
then we can get
\begin{equation}
\begin{split}
\hat{H}_1
&=\frac{J_1}{2}\sum_i\left(-S\hat{a}_i^{\dag}\hat{a}_i-S\hat{b}_{i+\delta_1}^{\dag}\hat{b}_{i+\delta_1}
+S\hat{a}_i\hat{b}_{i+\delta_1}^{\dag}+S\hat{a}_i^{\dag}\hat{b}_{i+\delta_1}+S^2\right)\\
&\ \ \ +\frac{J_2}{2}\sum_i\left(-S\hat{a}_i^{\dag}\hat{a}_i-S\hat{b}_{i+\delta_2}^{\dag}\hat{b}_{i+\delta_2}
+S\hat{a}_i\hat{b}_{i+\delta_2}^{\dag}+S\hat{a}_i^{\dag}\hat{b}_{i+\delta_2}+S^2\right)\\
&\ \ \ +\frac{J_3}{2}\sum_i\left(-S\hat{a}_i^{\dag}\hat{a}_i-S\hat{b}_{i+\delta_3}^{\dag}\hat{b}_{i+\delta_3}
+S\hat{a}_i\hat{b}_{i+\delta_3}^{\dag}+S\hat{a}_i^{\dag}\hat{b}_{i+\delta_3}+S^2\right)\\
&\ \ \ +\frac{J_1}{2}\sum_{j}(-S\hat{a}_{j-\delta_1}^{\dag}\hat{a}_{j-\delta_1}-S\hat{b}_{j}^{\dag}\hat{b}_{j}
+S\hat{b}_j^{\dag}\hat{a}_{j-\delta_1}+S\hat{b}_j\hat{a}_{j-\delta_1}^{\dag}+S^2)\\
&\ \ \ +\frac{J_2}{2}\sum_{j}(-S\hat{a}_{j-\delta_2}^{\dag}\hat{a}_{j-\delta_2}-S\hat{b}_{j}^{\dag}\hat{b}_{j}
+S\hat{b}_j^{\dag}\hat{a}_{j-\delta_2}+S\hat{b}_j\hat{a}_{j-\delta_2}^{\dag}+S^2)\\
&\ \ \ +\frac{J_3}{2}\sum_{j}(-S\hat{a}_{j-\delta_3}^{\dag}\hat{a}_{j-\delta_3}-S\hat{b}_{j}^{\dag}\hat{b}_{j}
+S\hat{b}_j^{\dag}\hat{a}_{j-\delta_3}+S\hat{b}_j\hat{a}_{j-\delta_3}^{\dag}+S^2),
\end{split}
\end{equation}
in which $J_1$, $J_2$ and $J_3$ represent nearest neighbor Heisenberg interactions in different directions. And we can get
\begin{equation}
\begin{split}
\hat{H}_2
&=D\sum_i(-1)\left[\frac{1}{2i}2S\left(\hat{a}_i^{\dag}\hat{a}_{i+a_2}-\hat{a}_i\hat{a}_{i+a_2}^{\dag}\right)
+\frac{1}{2i}2S\left(\hat{a}_{i+a_2}^{\dag}\hat{a}_{i+a_1}-\hat{a}_{i+a_2}\hat{a}_{i+a_1}^{\dag}\right)
+\frac{1}{2i}2S\left(\hat{a}_{i+a_1}^{\dag}\hat{a}_{i}-\hat{a}_{i+a_1}\hat{a}_{i}^{\dag}\right)\right]\\
&\ \ \ +D\sum_j\left[\frac{1}{2i}2S\left(\hat{b}_{j+\delta_2}^{\dag}\hat{b}_{j+\delta_1}-\hat{b}_{j+\delta_2}\hat{b}_{j+\delta_1}^{\dag}\right)
+\frac{1}{2i}2S\left(\hat{b}_{j+\delta_1}^{\dag}\hat{b}_{j+a_1+\delta_2}-\hat{b}_{j+\delta_1}\hat{b}_{j+a_1+\delta_2}^{\dag}\right)\right]\\
&\ \ \ +D\sum_j\frac{1}{2i}2S\left(\hat{b}_{j+a_1+\delta_2}^{\dag}\hat{b}_{j+\delta_2}-\hat{b}_{j+a_1+\delta_2}\hat{b}_{j+\delta_2}^{\dag}\right).\\
\end{split}
\end{equation}
\begin{equation}
\begin{split}
\hat{H}_3
&=g_J\mu_BB\sum_i(S-\hat{a}_i^{\dag}\hat{a}_i)-g_J\mu_BB\sum_j(S-\hat{b}_j^{\dag}\hat{b}_j),
\label{Zeeman}
\end{split}
\end{equation}
\begin{equation}
\begin{split}
\hat{H}_4=\mathcal{K}\sum_i(S^2-2S\hat{a}_i^{\dag}\hat{a}_i)+\mathcal{K}\sum_i(S^2-2S\hat{b}_i^{\dag}\hat{b}_i).
\end{split}
\end{equation}
In Eq. (\ref{Zeeman}), we apply a periodic magnetic field in the ferromagnetic Hexagonal lattice.
After taking the Fourier transform, we can find that the Hamiltonian can be expressed as
\begin{equation}
\begin{split}
\hat{H}_1
&=-3JS\sum_{\mathbf{k}}\left(\hat{a}_{\mathbf{k}}^{\dag}\hat{a}_{\mathbf{k}}+\hat{b}_{\mathbf{k}}^{\dag}\hat{b}_{\mathbf{k}}\right)
+3JS\sum_{\mathbf{k}}\gamma_{-\mathbf{k}}\hat{a}_{\mathbf{k}}\hat{b}_{\mathbf{k}}^{\dag}+3JS\sum_{\mathbf{k}}\gamma_{\mathbf{k}}\hat{a}_{\mathbf{k}}^{\dag}\hat{b}_{\mathbf{k}},
\end{split}
\end{equation}
in which $J=\frac{1}{3}\left(J_1+J_2+J_3\right)$, $\gamma_{\mathbf{k}}=\frac{1}{3J}\sum_{i}J_ie^{i\mathbf{k}\cdot\mathbf{\delta_{i}}}$,
\begin{equation}
\begin{split}
\hat{H}_2
&=\sum_{\mathbf{k}}\left[\triangle(\mathbf{k})\hat{a}_{\mathbf{k}}^{\dag}\hat{a}_{\mathbf{k}}-\triangle(\mathbf{k})\hat{b}_{\mathbf{k}}^{\dag}\hat{b}_{\mathbf{k}}\right],
\end{split}
\end{equation}
\begin{equation}
\begin{split}
\hat{H}_3
&=-g_J\mu_BB_{FM}\sum_{\mathbf{k}}(\hat{a}_{\mathbf{k}}^{\dag}\hat{a}_{\mathbf{k}}-\hat{b}_{\mathbf{k}}^{\dag}\hat{b}_{\mathbf{k}}),
\end{split}
\end{equation}
\begin{equation}
\begin{split}
\hat{H}_4
&=-2\mathcal{K}S\sum_{\mathbf{k}}(\hat{a}_{\mathbf{k}}^{\dag}\hat{a}_{\mathbf{k}}+\hat{b}_{\mathbf{k}}^{\dag}\hat{b}_{\mathbf{k}}),
\end{split}
\end{equation}
in which, $\triangle(\mathbf{k})=2SD\left[\sin{\frac{1}{2}(k_y-k_x\sqrt{3})}-\sin{k_y}+\sin{\frac{1}{2}(k_y+k_x\sqrt{3})}\right]$.

\end{widetext}


\begin{thebibliography}{99}
\bibitem{ref1}  M. Chang and Q. Niu, Phys. Rev. B 53, 7010 (1996).

\bibitem{ref2}  G. Sundaram and Q. Niu, Phys. Rev. B 59, 14915 (1999).

\bibitem{ref3}  D. Xiao, J. Shi, and Q. Niu, Phys. Rev. Lett. 95, 137204 (2005).

\bibitem{ref4}  D. Xiao, Y. Yao, Z. Fang, and Q. Niu, Phys. Rev. Lett. 97, 026603 (2006).

\bibitem{ref5}  Y. Gao, S. Yang, and Q. Niu, Phys. Rev. Lett. 112, 166601 (2014).

\bibitem{ref6}  Y. Gao and D. Xiao, Phys. Rev. B 98, 060402(R) (2018).

\bibitem{ref6.1} A. V. Chumak \textit{et al.}, Nature Phys 11, 453–461 (2015).

\bibitem{ref7} H. Katsura, N. Nagaosa, and P. A. Lee, Phys. Rev. Lett. 104, 066403 (2010).

\bibitem{ref8} Y. Onose \textit{et al.}., Science 329, 297 (2010).

\bibitem{ref9} M. Hirschberger, R. Chisnell, Y.S. Lee, and N.P. Ong, Phys. Rev. Lett. 115, 106603 (2015).

\bibitem{ref10} M. Hirschberger, J.W. Krizan, R.J. Cava, and N.P. Ong, Science 348, 106-109 (2015).

\bibitem{ref11} K. Tanabe, R. Matsumoto, J.-I. Ohe, S. Murakami, T. Moriyama, D. Chiba, K. Kobayashi and T. Ono, Phys. Status Solidi b 253, 783-787 (2016).

\bibitem{ref11.1} M. Hirschberger \textit{et al.}, Phys. Rev. Lett. 115, 106603 (2015).

\bibitem{ref12} R. Matsumoto and S. Murakami, Phys. Rev. B 84, 184406 (2011).

\bibitem{ref12.1} L. Zhang \textit{et al.}, Phys. Rev. B 87, 144101 (2013).

\bibitem{ref12.2} A. Rückriegel, A. Brataas, and R. A. Duine, Phys. Rev. B 97, 081106(R) (2018).

\bibitem{ref12.3} A. Mook, J. Henk, and I. Mertig, Phys. Rev. B 90, 024412 (2014).

\bibitem{ref12.4} A. Mook, J. Henk and I. Mertig, Phys. Rev. B 89, 134409 (2014).

\bibitem{ref12.5} T. Ideue \textit{et al.}, Phys. Rev. B 85, 134411 (2012).

\bibitem{ref13} K.A. van Hoogdalem, Phys. Rev. B 87, 024402 (2013) .

\bibitem{ref14} Z.-X. Li, Yunshan Cao, Peng Yan, Physics Reports 915,  1 (2021).

\bibitem{ref15} Hiroki Kondo and Yutaka Akagi, Phys. Rev. Research 4, 013186 (2022).

\bibitem{ref15.1} Rohit Mukherjee, Sonu Verma, and Arijit Kundu, Phys. Rev. B 107, 245426 (2023).

\bibitem{ref16} J. M. Luttinger, Phys. Rev. 135, A1505 (1964).

\bibitem{ref17} G. Tatara, Phys. Rev. Lett. 114, 196601 (2015).

\bibitem{ref19}  YuanDong Wang, Zhen-Gang Zhu and Gang Su, Phys. Rev. B 106, 035148 (2022).

\bibitem{ref18} T. B. Smith, L. Pullasseri, and A. Srivastava, Phys. Rev. Res. 4, 013217 (2022).



\bibitem{ref21}  Huiying Liu \textit{et al.}, Phys. Rev. Lett. 127, 277202 (2021).

\bibitem{ref21.01} L. Chotorlishvili \textit{et al.}, Phys. Rev. B 106, 014417 (2022).

\bibitem{ref21.1} S A Owerre, J. Phys.: Condens. Matter 30 245803 (2018).

\bibitem{ref22}  H. Varshney, R. Mukherjee, A. Kundu, and A. Agarwal, Phys. Rev. B 108, 165412 (2023).

\bibitem{Smrcka1977}  L. Smrcka and P. Streda, J. Phys. C 10, 2153 (1977).






\end{thebibliography}
\end{document}